\documentclass[journal]{IEEEtran}
\usepackage{cite}
\usepackage{amsmath,amssymb,amsfonts}
\usepackage{algorithmic}
\usepackage{graphicx}
\usepackage{textcomp}
\usepackage{setspace}
\usepackage{caption}
\usepackage[utf8]{inputenc} 
\usepackage{tabularx}
\usepackage{subfigure}
\usepackage{multirow}
\usepackage{adjustbox}
\usepackage{tabulary}
\usepackage{textcomp}
\graphicspath{ {./figures/} }
\def\BibTeX{{\rm B\kern-.05em{\sc i\kern-.025em b}\kern-.08em
		T\kern-.1667em\lower.7ex\hbox{E}\kern-.125emX}}
\usepackage{url}	
\begin{document}
\title{Digital Twin: Values, Challenges and Enablers}
\author{Adil~Rasheed$^{1,4}$,~Omer~San$^{2}$~and~Trond~Kvamsdal$^{3,4}$
\thanks{$^{1}$Department of Engineering Cybernetics, Norwegian University of Science and Technology, Trondheim, Norway (e-mail: adil.rasheed@ntnu.no)}
\thanks{$^{2}$Mechanical and Aerospace Engineering, Oklahoma State University, Stillwater, Oklahoma, 74078-5016 USA (e-mail: osan@okstate.edu)}
\thanks{$^{3}$Department of Mathematical Sciences, Norwegian University of Science and Technology, Trondheim, Norway (e-mail: trond.kvamsdal@ntnu.no)}
\thanks{$^{4}$Department of Mathematics and Cybernetics, SINTEF Digital, Trondheim, Norway}
}
\maketitle

\begin{abstract}
	A digital twin can be defined as an adaptive model of a complex physical system. Recent advances in computational pipelines, multiphysics solvers, artificial intelligence, big data cybernetics, data processing and management tools bring the promise of digital twins and their impact on society closer to reality. Digital twinning is now an important and emerging trend in many applications. Also referred to as a computational megamodel, device shadow, mirrored system, avatar or a synchronized virtual prototype, there can be no doubt that a digital twin plays a transformative role not only in how we design and operate cyber-physical intelligent systems, but also in how we advance the modularity of multi-disciplinary systems to tackle fundamental barriers not addressed by the current, evolutionary modeling practices. In this work, we review the recent status of methodologies and techniques related to the construction of digital twins. Our aim is to provide a detailed coverage of the current challenges and enabling technologies along with recommendations and reflections for various stakeholders.
\end{abstract}

\section{Introduction}
\label{sec:introduction}
With the recent wave of digitalization, the latest trend in every industry is to build systems and approaches that will help it not only during the conceptualization, prototyping, testing and design optimization phase but also during the operation phase with the ultimate aim to use them throughout the whole product life cycle and perhaps much beyond. While in the first phase, the importance of numerical simulation tools and lab-scale experiments is not deniable, in the operational phase, the potential of real-time availability of data is opening up new avenues for monitoring and improving operations throughout the life cycle of a product. Grieves~\cite{grievesDT} in his whitepaper named the presence of this virtual representation as "Digital Twin". 


Since its presence among the most promising technology trends in Gartner's recent report \cite{gartner2018}, the Digital Twin  concept has become more popular in both academia and industry (e.g., see \cite{iale} for a non-exhaustive glimpse at the major patented developments), apparently with many different attributes to include or exclude in its definition. For example, Hicks~\cite{hicks2019} differentiates a digital twin from a virtual prototype and redefines the concept as \textit{an appropriately synchronized body of useful information (structure, function, and behaviour) of a physical entity in virtual space, with flows of information that enable convergence between the physical and virtual states.} According to the authors in \cite{schluse2018experimentable}, \textit{digital twins represent real objects or subjects with their data, functions, and communication capabilities in the digital world.} 

Although the first mention of the word \textit{digital twin} can be traced back to only the year 2000 when Grieves mentioned it in the context of manufacturing, without a formal mention, several industries and organizations had been exploiting the idea at varying levels of sophistication. Some examples of these are the log of patients health information and history tracking, online operation monitoring of process plants, traffic and logistics management, dynamic data assimilation enabled weather forecasting, real-time monitoring systems to detect leakages in oil and water pipelines, and remote control and maintenance of satellites or space-stations.  
\begin{figure*}[h]
	\centering
	\includegraphics[width=\textwidth]{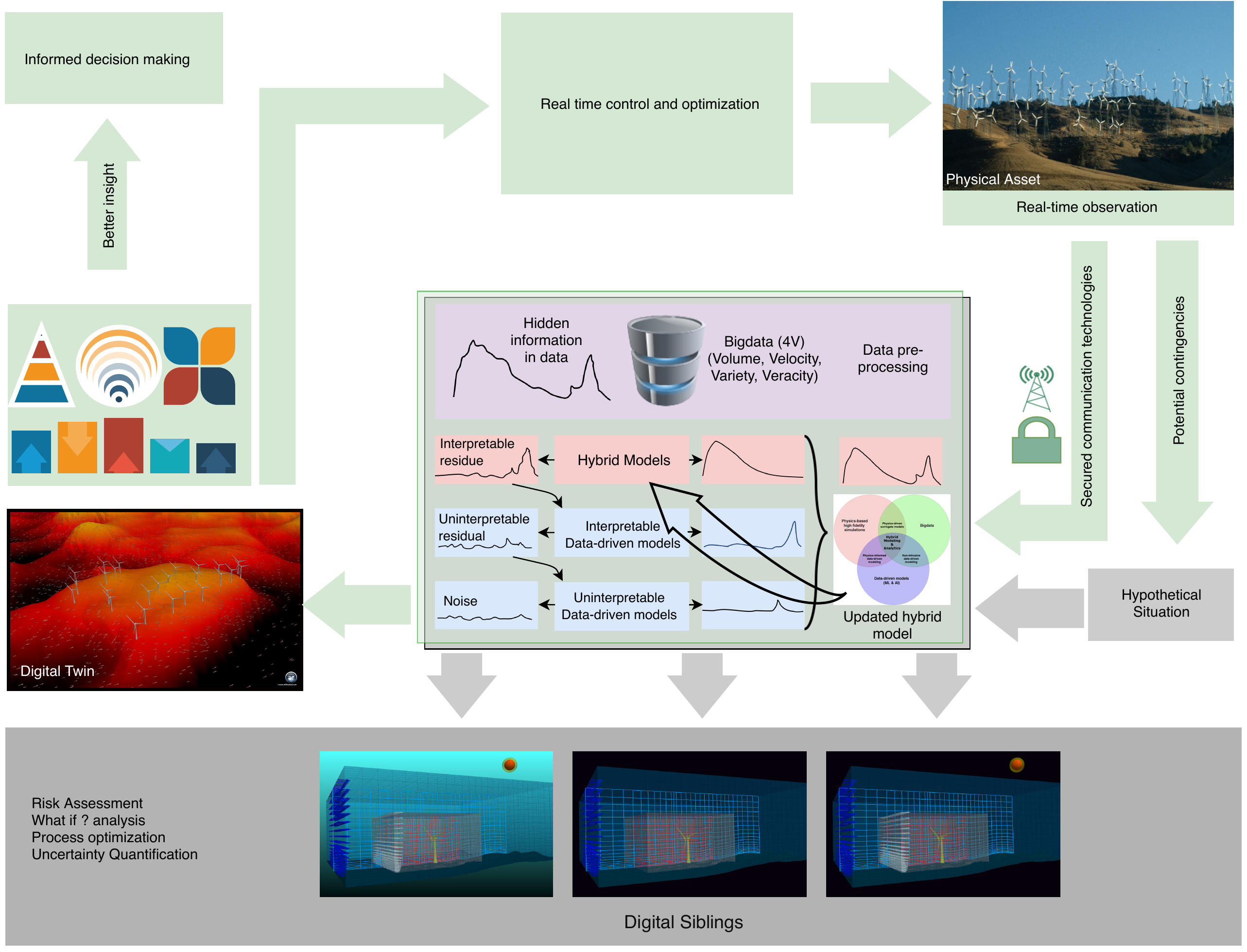}
	\caption{A generalized Digital Twin framework: Real-time data from an onshore wind farm is processed in real-time to infer the quantity of interest. This quantity can either be presented to decision makers or can be used for optimal control of the wind farm. The real-time data can also be perturbed to generate hypothetical scenarios that can be analyzed for ``what if'', risk analysis or evaluating the impact of mitigation strategies.  
	}
	\label{fig:conceptualDT1}
\end{figure*}
More recently, the necessity to formalize and utilize the full potential of digital twin concepts arises from a combination of technology push and market pull. While the need for online monitoring, flexibility in operation, better inventory management and personalization of services are the most obvious market pull, availability of cheap sensors and communication technologies, phenomenal success of Machine Learning (ML) and Artificial Intelligence (AI), in particular, Deep Learning (DL), new developments in the computational hardware (Graphics Processing Unit (GPU) and Tensor Processing Unit (TPU)), cloud and edge computing are certainly the major technology push. In this regard, it will not be an overstatement to say that the digital twin concept is going to bring revolution across several industry sectors. In this context, the current paper is a small effort to present a reasonably comprehensive overview of the concept. In doing so we plan to answer the following questions: 
\begin{itemize}
	\item What is a digital twin? 
	\item Is digital twin a new concept or has it existed with different names in the past?
	\item What values are expected out of the digital twin concepts?
	\item What are the major challenges in building digital twins?
	\item What are the existing and emerging enabling technologies to facilitate its adoption?
	\item What will be the social implications of this technology?
	\item What is expected from relevant stakeholders to extract maximum benefits from the digital twin technology?
\end{itemize}

In the current paper we view digital twin of a physical asset as the schematic shown in Fig. \ref{fig:conceptualDT1}. Any physical asset generates data which requires real-time processing (shown by green colored arrows and boxes) for mainly two purposes; informed decision making and for real-time optimization and control of the physical asset. Additionally, the recorded data can be perturbed to do offline "what-if ?" analysis, risk assessment and uncertainty quantification (shown in grey). In the latter context it is more appropriate to term the various virtual realizations of the physical asset as "Digital Siblings". With Fig. \ref{fig:conceptualDT1} in mind we start this paper with a brief description of eight values (Section \ref{secton:value}) that any digital twin is capable of generating followed by a presentation of the state-of-the-art technology in five diverse application areas (Section \ref{section:diverseapplication}). Common challenges across these application areas are identified in Section \ref{sec:challenges}.  To address these challenges, certain enabling technologies are required. Some of these technologies already exist but most of them are at different stages of development. A comprehensive overview of the enabling technologies is presented in Section \ref{sec:enabling} followed by a section on the socio-economic impacts of the technology (Section \ref{sec:socioimpact}). Finally, the paper concludes with reflection and recommendations targeted towards five distinct stakeholders.

\section{Value of Digital Twin}
\label{secton:value}
Building up on a report from Oracle \cite{oracleDT2017}, the following eight value additions of digital twin are identified:
\begin{enumerate}
	\item \textit{Real-time remote monitoring and control:} Generally, it is almost impossible to gain an in-depth view of a very large system physically in real-time. A digital twin owing to its very nature can be accessible anywhere. The performance of the system can not only be monitored but also controlled remotely using feedback mechanisms. 
	\item \textit{Greater efficiency and safety:} It is envisioned that digital twinning will enable greater autonomy with humans in the loop as and when required. This will ensure that the dangerous, dull and dirty jobs are allocated to robots with humans controlling them remotely. This way humans will be able to focus on more creative and innovative jobs. 
	\item \textit{Predictive maintenance and scheduling:} A comprehensive digital twinning will ensure that multiple sensors monitoring the physical assets will be generating big data in real-time. Through a smart analysis of data, faults in the system can be detected much in advance. This will enable better scheduling of maintenance.  
	\item \textit{Scenario and risk assessment:} A digital twin or to be more precise a digital siblings of the system will enable what-if analyses resulting in better risk assessment. It will be possible to perturb the system to synthesize unexpected scenarios and study the response of the system as well as the corresponding mitigation strategies. This kind of analysis without jeopardizing the real asset is only possible via a digital twin.
	\item \textit{Better intra- and inter-team synergy and collaboration:} With greater autonomy and all the information at a finger tip, teams can better utilize their time in improving synergies and collaborations leading to greater productivity. 
	\item \textit{More efficient and informed decision support system:} Availability of quantitative data and advanced analytics in real-time will assist in more informed and faster decision makings.
	\item \textit{Personalization of products and services:} With detailed historical requirements, preferences of various stakeholders and evolving market trends and competitions, the demand of customized products and services are bound to increase. A digital twin in the context of factories of the future will enable faster and smoother gear shifts to account for changing needs. 
	\item \textit{Better documentation and communication:} Readily available information in real-time combined with automated reporting will help keep stakeholders well informed thereby improving transparency. 
\end{enumerate}

\section{Diverse applications of digital twin}
\label{section:diverseapplication}
Five diverse applications of a digital twin are selected to get an in-depth understanding of the state-of-the-art, common challenges, corresponding solutions and future needs to bring more physical realism into it.   

\subsection{Health}
Health care is one of the sectors which is going to benefit the most from the concept of digital twin. The emergence of smart wearable devices, organized storage of health data of individuals and societies, need for personalized and targeted medication and inter-weaving of engineering and medical disciplines are the major driving forces for the realm of smart and connected health. 

Psychologists have begun to use physical activity levels using actigraphs to predict the onset of different episodes of bipolar disorder \cite{Krane-Gartiser2018actigraphy}. Fern\'{a}ndez-Ruiz~\cite{Fernandez-Ruiz2018} suggests combining computational simulations with tissue engineering for more reliable, successful and predictable clinical outcomes. In \cite{laaki2019prototyping}, the authors give insight into how a digital twin framework for remote surgery might look like. Bruynseels et al.~\cite{bruynseels2018digital} projects digital twin as an emerging technology building silico representations of an individual that dynamically reflects molecular status, physiological status and life style over time. Zhang et al.~\cite{Zhang2018Organ} gives an overview of the latest developments in organ-on-a-chip (OOC) engineering. OOC synthesized from the tissues of a patient offers an alternative to conventional preclinical models for drug screening. The chip can replicate key aspects of human physiology crucial for the understanding of drug effects. By combining several such OOC one can generate a full body-on-a-chip (BOC). Skardal et al.~\cite{skardal2016organoid} gives the progress made in this direction. Although both OOC and BOC are seen as disruptive technologies in the context of digital twin (or digital sibling to be more precise), in the more near future a digital twin of human body can be worked out using most of the technologies and data already available. To begin with, the health care system in developed countries already has recorded health history of its inhabitants. However, one issue is that this monitoring is not regular and a big chunk of the young population is left out because they visit physicians only when sick. To this end, the data collected by smart wearables might hold huge value since it can help balance the database enabling longitudinal studies. Furthermore, in the presence of a digital twin of the human body of a patient, a surgeon can already train himself even before physically conducting the surgery. However, for the digital twin to be a reality there are many technical, ethical and privacy issues that should be resolved. Physical models for biochemistry, fluid and nutrient transport, and other mechanistic models are required to better simulate the internal working of a human body (organs, blood flow, signal transmission etc). The authors in \cite{oran2009computational} explained a "computational man" vision, which is a dynamic fast-running computational model of major physiological systems, such as the circulatory, respiratory or excretory system in a human body. In the prototype for remote surgery \cite{laaki2019prototyping}, the authors highlighted the importance of low latency, high level of security and reliability hinting at how important role communication technologies (like 5G) and cybersecurity (encryption technologies) are going to play. Efficient Computer Aided Modeling (CAM) and Virtual Reality (VR) will also be instrumental in this context. Jimenez et al.~\cite{jimenez2020health} also advocated building digital twin technologies in medicine (see, for example, \cite{lee2011challenges,dey2018medical} for a detailed description of medical cyber-physical systems). Kocabas et al.~\cite{kocabas2016emerging} detailed such medical cyber-physical system considering the layers of data acquisition, data preprocessing, cloud system and action module. The secure integration of wireless body area networks (WBAN) with cloud computing platforms and Internet of Things (IoT) networks has been considered as another major challenge in healtcare applications \cite{latre2011survey}.       

\subsection{Meteorology} 
The meteorological institutes all over the world have been using the concepts of digital twin extensively. They utilize solid models of a region (terrain, buildings), high fidelity physics simulators and big data coming from multiple sources to provide both long and short term weather prediction which are ultimately communicated using stunning rendering of the numerical results via web browser, television or mobile applications (see, for example, \cite{lazo2011us,bauer2015quiet,ronda2017urban}). The big data (terrain data, satellites, masts, radar, radiosonde, sea buoy, drones, LIDAR) handled in meteorology is characterized by the high 4Vs (volume, velocity, variety and veracity). In this application area, one can find some of the most advanced methods for big data assimilation in massively parallelized numerical simulators capable of operating in real-time, and excellent long-term and well documented data archival and retrieval mechanisms. 

One of the reasons for the immense success of the digital twin concept within the field of meteorology is a relatively much lower bar regarding privacy and confidentiality of the data. There is a lot to be learned from this field. While it will not be an exaggeration to term weather forecasting as the most matured digital twin concept, there is still enough scope for improvements with relative ease. The meteorological services have poor quality in those locations throughout the world where there are not enough weather stations or observation data. The precision meteorology aims to improve weather forecasting by updating predictions through numerical modeling, data collection, data analytics, and technology adoption enhancement. A greater than ever penetration of smartphones in almost all societies has remained an underutilized technology. By 2020 there will be more than 6 billion smartphones in the world. Compared to the paltry 10,000 official weather stations this number is huge. Each smartphone is equipped to measure raw data such as atmospheric pressure, temperature and humidity to access atmospheric conditions. While the analysis of data from only a few phones might not yield much, analysis of data from billions of phones will certainly be game changer. In fact Price et al.~\cite{price2018smartphones} has already demonstrated the use of smartphones for monitoring atmospheric tides. Nipen et al.~\cite{nipen2019private} has recently shown the value of crowd sourced data in improving operational weather forecasts.

In weather and climate centers, large volume and large variety observational data streams have been used to improve our knowledge of the current state of the atmosphere and ocean, that provide an initial condition for forecasts of future states. As an integrated approach, observing system simulation experiment (OSSE) systems mimic the procedures used to analyze such satellite, balloon, or surface tower observations for specifying states of the atmosphere \cite{wang2008hybrid,errico2013development,prive2013validation,benjamin2016north,james2017observation}. An OSSE can be considered as a digital twin to use computational models to test new systems before their physical prototype is actually built or deployed. It is anticipated that in the future such OSSE techniques will be applied to diverse domains with the use of increasingly advanced and sophisticated simulations of nature and observations \cite{hoffman2016future}.

A near real-time weather forecast is a must for environmental situational awareness for improved decision-making scenarios \cite{benjamin2016north}. Examples include, but are not limited to, air traffic management, severe-weather watch, wind-turbines adaptation, and so on. To achieve so, an increasing number of observations should be assimilated instantaneously since the lifespan of many weather phenomena can be less than a few minutes. Solving a purely physics based model with the required resolution while being compatible with the time-scales of natural phenomena might be prohibitive. Therefore, a parallel accelerated digital twin, capable of duplicating the physical system with acceptable accuracy, could offer a realistic solution to assimilate these short-time observations. The role of this twin is to give quick guidelines whether or not a decision needs to be taken. This is specifically important with the intrinsic stochastic characteristics of weather systems, and hence the model should be able to monitor those seemingly-random sequential observations \cite{dee1998data,dee2000data,chinesta2018virtual}. Data-driven tools do an excellent job in terms of online time-cost, however relying only on data and disregarding the known and well-established physics is not a wise step. Hybridization techniques have been demonstrated to give superior performance to either individual components. Therefore, a digital assimilation twin can be built to provide nature-informed indicators for near-time events.

Besides, to test new models and assimilation tools before adopting them into current operational platforms, they need to be tested with some synthetic observations simulated from realistic nature runs (NR) \cite{hoffman2016future}. To maintain a realistic benchmark, these NR should mimic the natural phenomena with their respective time-scales and provide timely observations. ML and AI tools have shown extraordinary performance learning from available data. Supplying a digital twin with sufficient amount of real (noisy) data and posing the well-established physical constraints can provide a trained and informed nature replica with realistic stochastic behavior. In particular, this stochasticity is important for testing the predictability of extreme events that usually occur without  previous long-term indicators \cite{easterling2000observed, cousins2014quantification,joo2017extreme}.

\subsection{Manufacturing and process technology}
Manufacturing and process technology industries have a history of exploiting the digital twin concepts without explicitly mentioning it. However, with the current trend of digitalization and demand for customized, high quality products in highly variable batches with short delivery times, the industries are forced to adapt their production and manufacturing style. In this regard, even these industries are reevaluating the digital twin concepts and are trying to include latest enabling technologies (discussed later in Section~\ref{sec:enabling}) in their workflow. According to a report by BCG~\cite{bcg}, Industry 4.0 is seen as the convergence of nine digital technologies: big data and analytics, cybersecurity, industrial IoT, simulation, augmented reality, additive manufacturing, advanced robotics, cloud services, horizontal and vertical system integration. Borangiu et al. \cite{borangiu2019DTmanufacturing} stresses the importance of cloud services in the context of resource virtualization as an enabler for Factory of the Future (FoF) known by different initiatives Industry 4.0 (Germany), Advanced Manufacturing (USA), e-factory (Japan) and Intelligent Manufacturing (China) but similar goals.  

Mukherjee and DebRoy~\cite{mukherjee2019adigitaltwin} and Knapp et al.~\cite{knapp2017buildingblockDTAM} proposed a conceptual framework of a digital twin for 3D printing. In their models, the computationally demanding mechanistic models (for heat transfer, fluid flow and material deposition and transformation) are certainly the bottleneck. By providing an overview of the historical perspective on the data lifecycle in manufacturing, in \cite{tao2018DTbigdata} and \cite{tao2018DTproduction} the authors highlighted the role of big data in supporting smart manufacturing. The state-of-the-art of digital twin applications in industrial settings has been studied systematically in \cite{tao2018digital}, where the authors concluded that the most popular digital twin application area is developing the advanced prognostics and health management (PHM) systems. A five-level cyber-physical system structure has been proposed specifically for PHM in manufacturing applications \cite{lee2015cyber}. In determining the major concepts and key methodologies, the authors in \cite{zhong2017intelligent} reviewed intelligent manufacturing systems.

A functional mock-up interface has been offered as a tool independent standard to improve model-based design between the manufacturers and suppliers \cite{blochwitz2011functional}. Such a platform independent co-simulation framework can be found in \cite{hatledal2019language}. In \cite{negri2019fmu}, the authors detailed the use of functional mock-up units for the digital twin applications. A digital twin approach has been also highlighted in \cite{soderberg2017toward} as a real-time control and optimization tool towards individualized production systems.

\subsection{Education}
Massive Open Online Courses (MOOCs), which make university courses available for free or at a very nominal cost is a disruptive technology challenging traditional classroom teaching. In this paradigm change in pedagogy, we are undergoing a plethora of opportunities in personalizing education for a more effective teaching. Different students based on their natural talent, interest and backgrounds have different needs. Although universities and schools are recording more and more data about the students, the feedback loop is not complete. A digital twin for students will help counter this shortfall. With the emergence of artificial intelligence driving bots, personalized education and counseling will be more and more available at no additional cost. In fact, a brief discussion on the power of artificial intelligence-powered personalization in MOOC learning can be found in \cite{yu2017AIpoweredMOOC}. A personalized adaptive learning frameworks has been also constructed in \cite{peng2019personalized} using a smart learning environment. In \cite{mitrofanova2019modeling}, the authors developed a smart education concept incorporating data mining tools. The role of IoT in constructing a smart educational process can be found in \cite{abdel2019internet}. The authors in \cite{adu2014smart} explained the gradual progress from e-learning towards to m-learning (mobile learning), u-learning (ubiquitous learning), and s-learning (smart learning). In \cite{sabir2018an}, the authors demonstrated a digital data management framework for accreditation purposes in order to help assessment process considering data collection, characterization, analysis and implementation. This might also help to map program's educational objectives to student outcomes \cite{osman2018benchmark}.  The authors in \cite{assante2019smart} discussed a set of demanding skills that are critically important for the emerging digital twin paradigm and highlighted relevant business model changes to accommodate with the Industry 4.0 digital revolution. The growing gap between industry and academia has been highlighted in \cite{berman2018realizing}, where the authors drew attention to tremendous opportunities from a data science perspective. Combining the digital twin of the education sector with that of other industrial sectors, it would be more efficient in grooming targeted talents.

\subsection{Cities, Transportation and Energy Sector} 
The digital twin relevant technologies become more mature to offer smart solutions in construction, transportation and energy sectors. In \cite{kent2019early}, the authors studied a digital twin concept to engage citizens with city planning with an idea of pairing physical objects with their digital counterparts. The authors in \cite{hao2015scalable} discussed the requirements of the scalable cloud model that can be used for smart cities. A collection of essays about Singapore, arguably considered the smartest city in the world, can be found in \cite{kiang201650}. In his editorial note, Batty~\cite{batty2018digital} discussed the progress towards digital twins concerning urban planning and city science.  A smart city paradigm that can enable increased visibility into cities' human-infrastructure-technology interactions has been explained in \cite{mohammadi2017smart}, in which spatio-temporal city dynamics was integrated into a data analytics platform at the real-time intersection of reality and virtuality.

Smart grids use sensors and digital communications technologies to detect and react to local changes in power usage. A cyber-physical system perspective to the smart grid technologies can be found in \cite{yu2016smart}. The adaptation of the digital twin technology for applications in power system control centers has been recently studied in \cite{brosinsky2018recent}. In \cite{joseph2018predictive}, the authors used a faster than real-time digital twin in predicting dynamic behaviour of smart grid systems.

In \cite{besselink2016cyber}, the authors outlined a cyber-physical system approach to control large scale road freight transportation problems including minimized fuel consumption with integrated routing and transport planning. An integration approach between vehicles and mobile cloud computing has been discussed in \cite{wan2014vcmia}. A vision for the integration of cyberspace and physical space has been discussed in \cite{sampigethaya2013aviation} to tackle some challenges in aviation industry. In \cite{vinot2014tangible}, the authors provided a hybrid artifact solution for air traffic control, in which paper and digital media are identical and have equal importance, by suggesting tangible interaction and augmented reality. A discussion on digital artifacts related to simple examples of structural engineering can be found in \cite{chacon2017physical}. 

The recent technologies related to the Industry 4.0 have been recently reviewed for oil and gas sector \cite{lu2019oil}. Highlighting the digital shift, the authors in \cite{mohamed2019leveraging} discussed the integration of such technologies for improving energy efficiency in smart factories with the aim of both reducing both production costs and greenhouse gas emissions. Fault detection and diagnostics issues in built environment can be found in \cite{dong2014bim} towards improved Building Information Modeling (BIM) systems. 
In \cite{alonso2019sphere}, the authors introduced BIM digital twins for better assessment and development of the design, construction and performance of residential buildings. A full scale digital twin of a district heating and cooling network aligned with the visions of Industry 4.0 can be found in \cite{loureiro2018district}.

\section{Common Challenges} \label{sec:challenges}
\begin{figure}[h]
	\centering
	\includegraphics[width=1.1\linewidth]{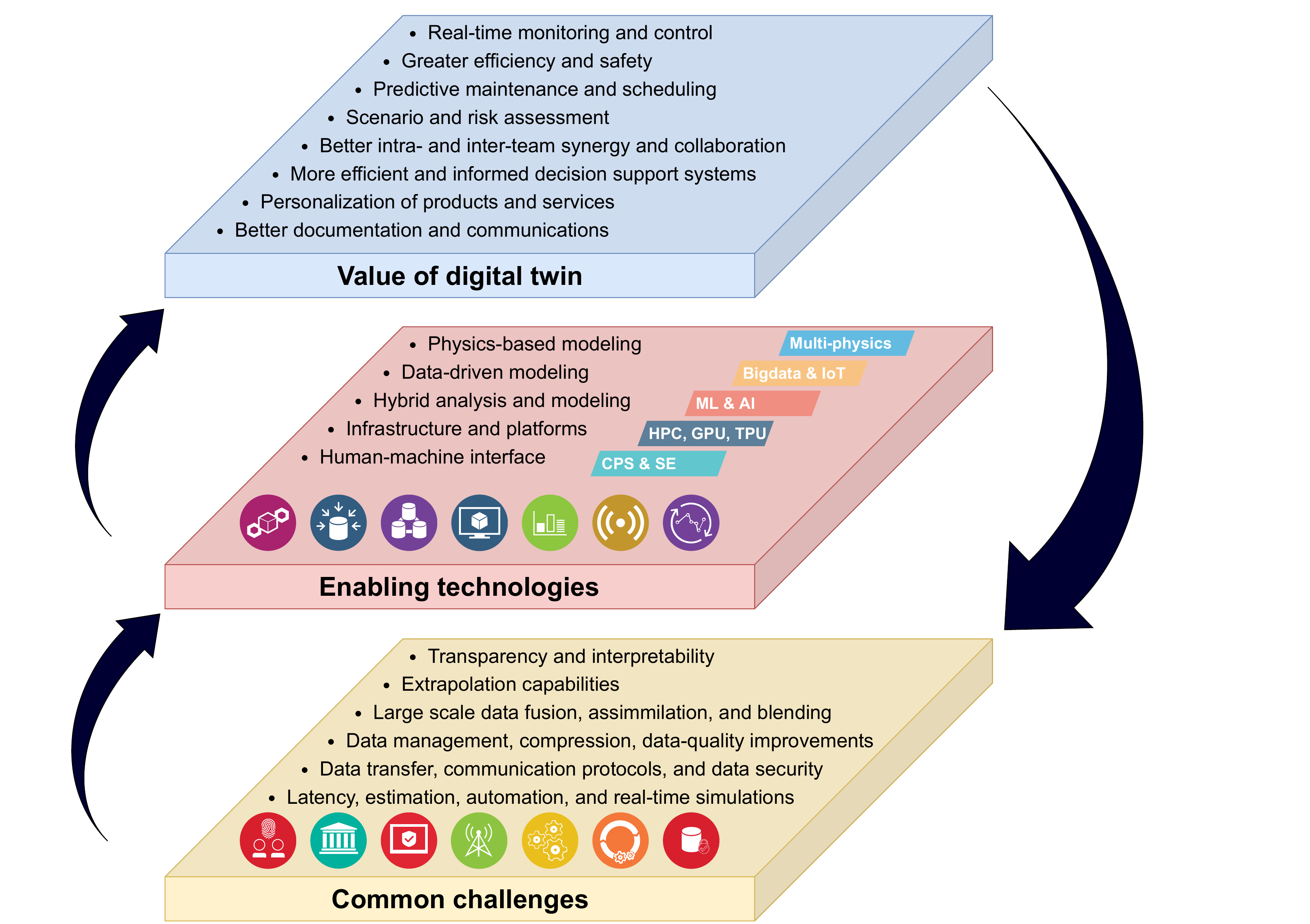}
	\caption{Interaction between value generation, common challenges and enabling technologies}
	\label{fig:schematic}
\end{figure}

As seen in the previous section, the concept of digital twin is already in use in several application areas, however, in order to make digital twins completely indistinguishable from their physical counterparts and a boon, many challenges have to be addressed. Fig.~\ref{fig:schematic} summarizes the interactions between value generation, common challenges and enabling technologies. In a similar context \cite{gunes2014survey}, the authors provided a compound table that touches on the importance of major cyber-physical system challenges for various application domains. Their discussion includes detailed definition for interoperability (composability, scalability, heterogeneity), security (integrity, confidentiality, availability), dependability (reliability, maintainability, availability, safety), sustainability (adaptability, resilience, reconfigurability, efficiency), reliability (robustness, predictability, maintainability), and predictability (accuracy, compositionality). In \cite{he2018surveillance}, the authors discussed various signal processing related challenges in digital twin technologies. Strict requirements for explainability and the resulting data protection laws now require that the decisions that are made for human should be explainable to humans \cite{eureg}. For example, human factor perspectives on auto driving systems can be found in a recent study \cite{kyriakidis2019human}. For example, with the rapid development of the AI technologies in mind, the moral machine experiment \cite{awad2018moral} provides many insights how machines might make moral decisions. 

From a purely technical perspective, data security and management, data quality improvements, latency, real-time simulations, large scale data fusion and assimilation, intelligent data analytics, predictive capacity, transparency and generalization of technologies across diverse application areas are considered main challenges in developing digital twin technologies. In the next section we focus on enabling technologies that can have maximum impact on addressing these challenges irrespective of the application areas.
\section{Enabling Technologies} \label{sec:enabling}
It is quiet evident from the preceding section on common challenges that we need to develop technologies that will address those challenges. In this section we attempt to cover the enabling technologies under five major categories: physics-based modeling, data-driven modeling, big data cybernetics, infrastructure and platforms, and human-machine interface. 
\subsection{Physics-based modeling}
\begin{figure}[htbp]
	\centering
	\includegraphics[width=\linewidth]{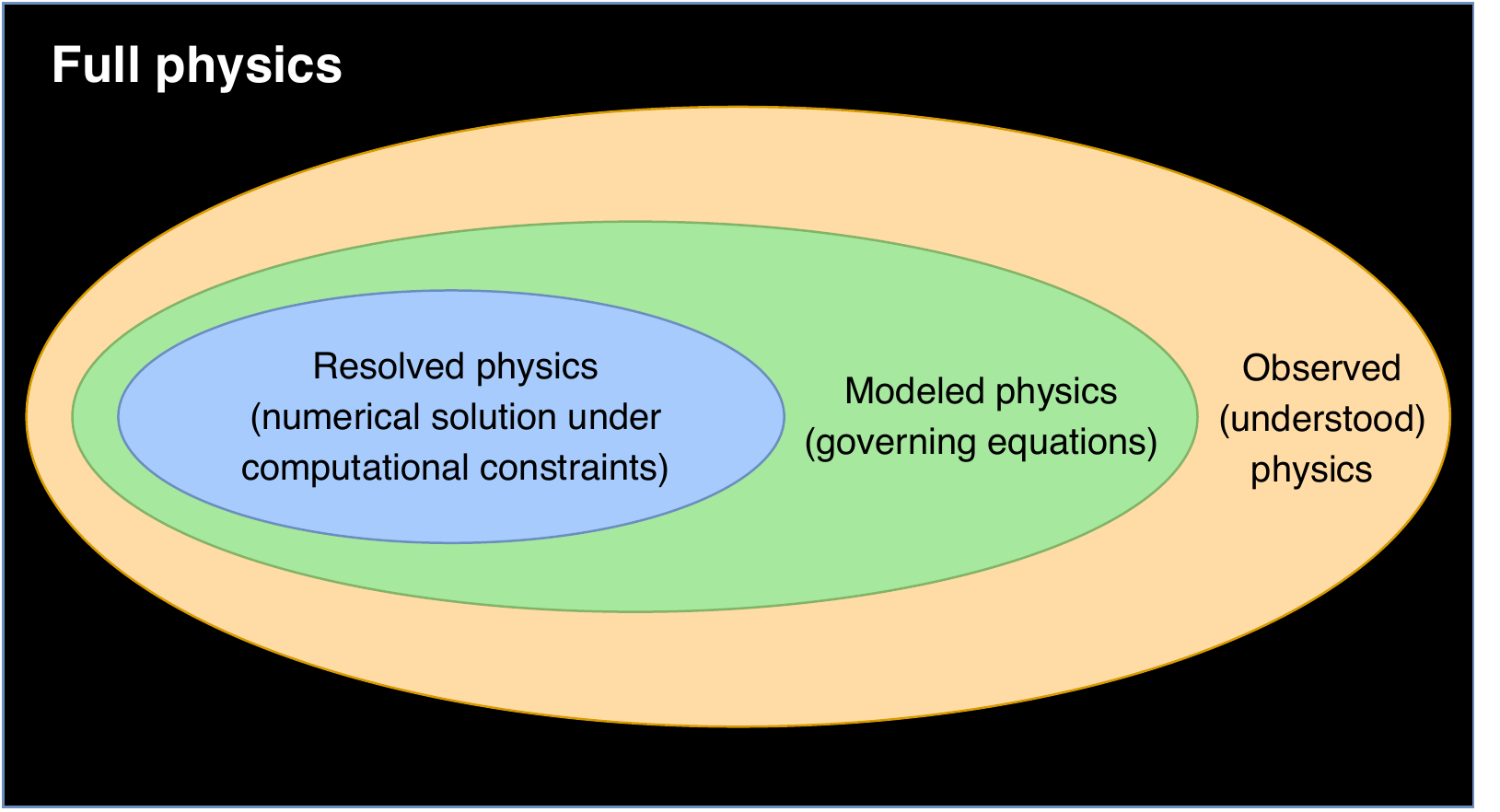}
	\caption{Physics based modeling: Based on first principles but can only model part of known physics due to assumptions at different stages}
	\label{fig:pbm}
\end{figure}
So far, the engineering community has been driven mostly by a physics-based modeling approach (e.g., see Fig.~\ref{fig:pbm} for broader hierarchical stages). This approach consists of observing a physical phenomenon of interest, developing a partial understanding of it, putting the understanding in the form of mathematical equations and ultimately solving them. Due to partial understanding and numerous assumptions along the line from observation of a phenomena to solution of the equations, one ends up ignoring a big chunk of the physics. The physics based approach can be subdivided broadly into experimental and numerical modeling. 

\subsubsection{Experimental Modeling}
This approach consists of doing a lab or full scale experiment or surveys to understand a process or a phenomena, developing correlations or models of quantities that can not be directly measured (or is expensive to measure) that can later be used in the the context of a digital twin. Hunt experiment \cite{hunt4} to collect health data, Bubble experiment \cite{bubble} to collect urban boundary layer measurements, wind tunnel experiments to understand the physics of fluid flows are some of the examples that have been used to develop simplified parameterizations which are now widely used in the respective fields. Laboratory scale experiments are conducted on reduced scale of the problem under investigation. While it is possible to work with a reduced scale model, the scaling laws based on physics are almost always impossible to satisfy and hence limit their utility. Furthermore, both lab scale as well as field experiments can be extremely expensive, and they are therefore conducted for very coarse spatio-temporal resolutions.  

\subsubsection{Three-dimensional (3D) Modeling} 
The first step towards numerical modeling is 3D modeling. It is the process of developing a mathematical representation of any surface of an object. In most of the digital twins, 3D models are the starting point. The 3D models can either be constructed by a 3D scan of the object or through specialized software using equations and finally represented in terms of curves and surfaces. Almost all 3D models can be categorized into solid or shell models. In the context of digital twin, the 3D model goes as input to the physical simulators and hence its quality is of utmost importance. In this regards the representation of 3D finite element models using splines~\cite{Hughes2005iac} (denoted isogeometric analysis) is interesting since it enables streamlining the whole workflow from 3D modeling to simulation and visualization of results, see~\cite{Stahl2017ppa}.

During the last decade a great progress towards adaptive refinement of isogeometric 3D models has been achieved. One popular method for local refinement is the Locally Refined B-splines (LR B-Splines), see~\cite{Dokken2013pso} and ~\cite{Johannessen2014iau}. In~\cite{skytt2015locally} and~\cite{dokken2018trivariate} LR B-Splines have been proposed for achieving more compact representation of the object geometry. In fact LR B-Splines approach appears to be very efficient in achieving high data compression ratios, which is very advantageous in a digital twin context. These new developments facilitate real-time geometry modifications (e.g., in simulating surgeries, deformation of structures) and associated simulations (e.g., fluid-structure simulations).

\subsubsection{High Fidelity Numerical Simulators}
In order to add physical realism to any digital twin, the governing equations derived through physical modeling need to be solved. For simplified equations, sometimes analytical solutions are possible to derive but most of the time due to complexities, the equations need to be solved numerically on computers. Various discretization techniques over time have been developed for this. Some of the commonly used methods belong to one of the following categories:  Finite Difference Method (FDM), Finite Element Method (FEM), Finite Volume Method (FVM) and Discrete Element Method (DEM). A detailed history of these methods can be found in~\cite{chung2010computational}. They have been extensively used in many open-source and commercial multi-physics simulation packages (e.g., OpenFOAM, FEniCS, Elmer, ANSYS, Comsol, STAR-CCM+ etc). 

A great advantage of any physics based modeling approaches is that they are generally less biased than data-driven models since they are governed by the laws of nature. However, the choices of which governing equation (e.g. turbulence model in fluid flows) should be applied for a given physical system might be biased in the sense that different scientists/engineers have different preferences (e.g. due to different educational background), but this kind of bias is transparent as long as the governing equations are stated. Furthermore, physics based models are highly interpretable along with their generalizability to seemingly very different problems governed by the same physics. At the same time, however, these models can be prone to numerical instability, can be too computationally demanding, can have huge errors owing to uncertainty in modeling and inputs, and lack of robust mechanisms to assimilate long term historical data. Another problem associated with numerical modeling of man made objects is the incompatibility between the way 3D geometries are modeled in CAD-systems and the way the equations are solved in numerical simulators. It is estimated that 80\% of the total time goes into cleaning the geometries and pre-processing, if not more, and only the remaining 20\% for actual simulation and analysis. To alleviate this problem Hughes et al.~\cite{Hughes2005iac} in 2005 came up with a novel method for analysis known as the Isogemetric Analysis. The approach offers the possibility of integrating finite element analysis with any conventional NURBS-based CAD design tools. Since 2005, there has been an explosion of publication in this field. The method has been utilized to solve a large array of problems ranging from fluid mechanics \cite{nordanger2016implementation}, \cite{nordanger2015simulation}, to solid mechanics to fluid-structure interaction \cite{nordanger2016numerical}, \cite{bazilevs3dsimulation2} as well as adaptive computational mechanics based on a posteriori error estimates~\cite{Kumar2015asa} and ~\cite{Kumar2017spr}. 

Today advanced numerical simulators are e.g. used to improve surgical procedures and develop patient-specific models that enable predictive treatment of cardiovascular diseases. High-performing simulators capable of handling a billion degrees of freedom are opening new vistas in simulation-based science and engineering and combined with multiscale modeling techniques have improved significantly the predictive capabilities. Thus, the dream of establishing numerical laboratories (e.g. numerical wind tunnels) can now be realized for many interesting real world systems. Despite their immense success, the use of high-fidelity simulator has so far been limited to the design phase of man made engineering objects/systems. Unless their computational efficiency is improved by several orders of magnitude their full potential will remain under-utilized in a digital twin context. 
However, the great advances of the high-performance simulators during the last two decades qualify (many of) them to be denoted "high-fidelity" models that can serve to develop "Reduced Order Models" (ROM), see below,  which may be used to establish predictive digital twins.


\subsection{Data-driven modeling}
\begin{figure}[htbp]
	\centering
	\includegraphics[width=\linewidth]{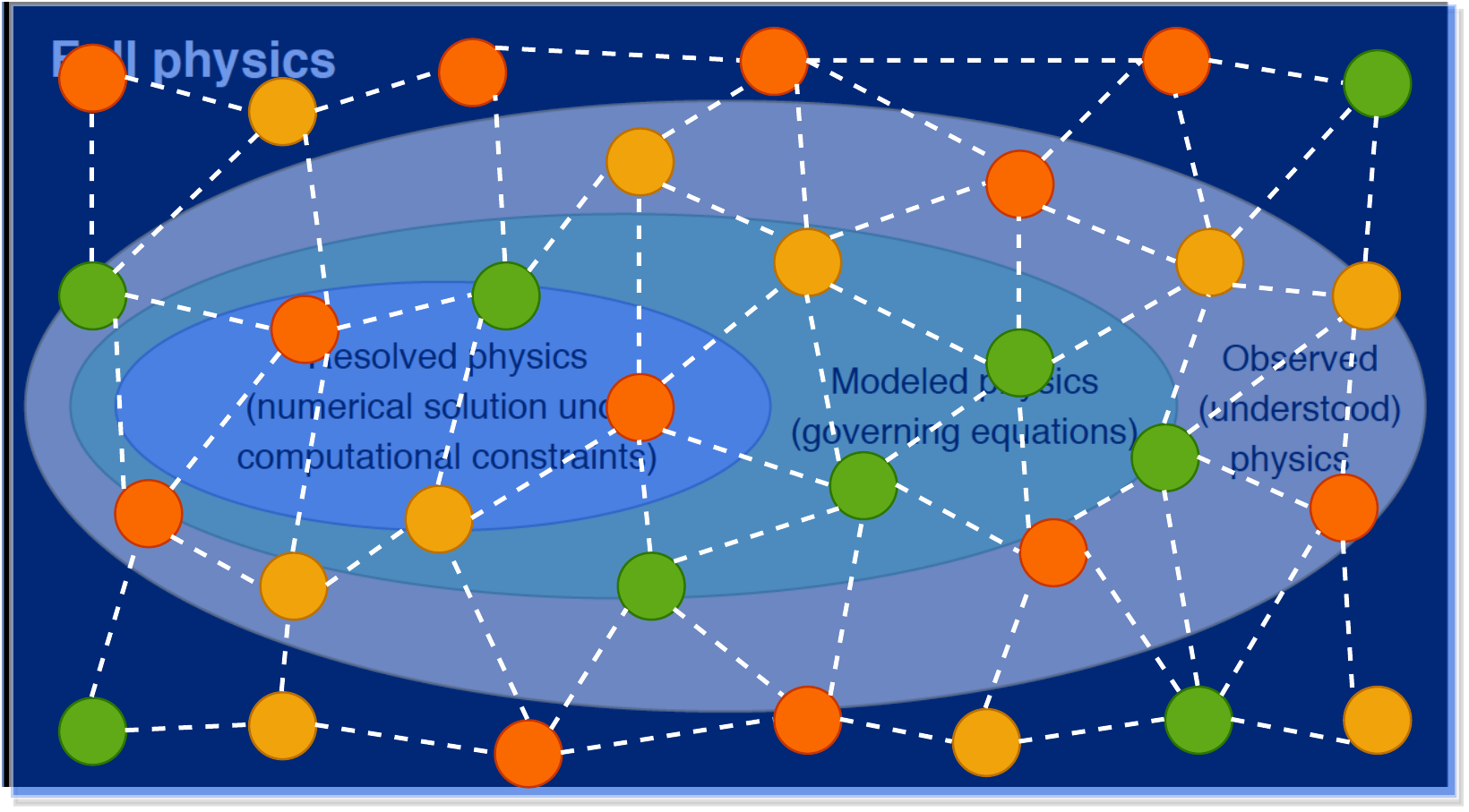}
	\caption{Data-driven modeling: Data is seen as a manifestation of both known and unknown physics}
	\label{fig:ddm}
\end{figure}

While physics based models are the workhorse at the design phase, with the abundant supply of data in a digital twin context, opensource cutting edge and easy-to-use libraries (tensorflow, torch, openAI), cheap computational infrastructure (CPU, GPU and TPU) and high quality, readily available training resources, data-driven modeling is becoming very popular. Compared to the physics based modeling approach, as illustrated in Fig.~\ref{fig:ddm}, this approach is based on the assumption that since data is a manifestation of both known and unknown physics, by developing a data-driven model, one can account for the full physics. In this section we present the state-of-the-art starting from data generation and safety to advanced data driven modeling techniques. 
\subsubsection{Data Generation}
One of the main reasons for the recent popularity of digital twins is the availability of cheap and miniaturized sensors which are capable of recording all possible kinds of imaginable data: text, audio, RGB images, hyperspectral images and videos. These miniaturized sensors can be installed on moving vehicles, drones, small satellites and underwater vehicles to quickly span all the four dimensions (three in space and one in time). Crowd sourcing of data because of the popularity of smart phones is also contributing massively towards data generation. It has been used to build up comprehensive databases of urban landcover World Urban Database and Access Portal Tools (WUDAPT \cite{wudapt}) database and local weather using NETATMO stations \cite{netatmo}. Online surveys and activities on social networking sites have created huge databases which are being used for training advanced ML models that are now being widely used for analyzing data.        

\subsubsection{Data Preprocessing, Management and Ownership}
With huge amount of data comes the problem of quality, management and ownership resolution of the data. For a correct working of digital twin there will be a need to improve the quality of data and compress them on the fly. Algorithms are required for automatic outlier detection and filling of missing data. At the moment very simplistic models like Principal Component and multidimensional interpolation techniques are utilized to fix these issues. However, more advanced methods like Restricted Boltzman Machine and Generative Adversarial Networks can be employed. Martens \cite{martens2015quantitative} has developed IDLE (Intensity observed =  Displacement model of (Local intensity model) + Error) methods for on the fly learning from big data stream. As mentioned earlier LRB-S representations can be used for achieving massive compression ratios. The problem of ownership will be a tough nut to crack. The issue of data ownership can be complicated due to the involvement of different stakeholders who participated in the generation, processing, value addition or analysis of data. This can jeopardize the smooth working of any digital twin concept. 

\subsubsection{Data Privacy and Ethical Issues}
For successful implementation of digital twins, trust in information system is extremely critical especially when stringent privacy laws are getting shaped, digital twin will be used more and more in safety critical applications. In this context, Nakamoto~\cite{nakamoto2008} presented two game changing concepts. The first one is bitcoins and the other is blockchain. Blockchain is a distributed, secured and verifiable record of information (e.g., data, transactions) linked together in a single chain. The information is stored inside cryprographic blocks connected in a sequential order in a chain. The validity of each new block is verified by a peer-to-peer network before it is linked to the blockchain using a crytographic hash generated from the contents of the previous block. This ensures that the chain is never broken and each block is permanently recorded. This provides maximum security, traceability and transparency in applications where blockchains can be employed. In the context of digital twin where traceability and security of information is of utmost importance, blockchain is foreseen to play an important role. Mandolla et al.~\cite{mandolla2019building} gives a nice insight into the exploitation of blockchain technology in building a digital twin for additive manufacturing with focus on aircraft industry, and Li et al.~\cite{li2019blockchain} introduces an extended socio-technical framework for the implementation of blockchain into the construction industry. The authors in \cite{chen2019blockchain} propose a blockchain based searchable encryption for electronic health record sharing. Reyna et al.~\cite{reyna2018onblockchain} gives a detailed overview of integration of the blockchain technology with IoT, its challenges (related to scalability, privacy, security and a total dearth of censorship) and latest research work to alleviate the associated shortcomings. 

\subsubsection{Machine Learning and Artificial Intelligence}
Smart data analysis using ML and AI is expected to play a major role in the context of digital twin. ML is the scientific study of algorithms and statistical models that computer systems use in order to perform tasks effectively without being explicitly programmed to do so and instead relies on learning from data alone. Although the term was coined in 1959 by Arthur Samuel, an American pioneer in the field of computer gaming and artificial intelligence, its real potential is being realized only recently when computers have begun to outperform humans in even creative tasks like art generation, script writing, text summarization, language translation and language interpretation. Any ML can be broadly categorized into supervised, unsupervised and reinforcement learning. 

Supervised learning is aimed at learning a mapping from independent variables to dependent variables. The mapping tasks can be either regression or classification. Most of the commonly used supervised algorithms are Linear Regression, Logistic Regression, Support Vector Machine, Decision Trees, Random Forest and Artificial Neural Networks. The recent success of ML in outperforming humans in image classification \cite{He2015}, \cite{inception-4}, \cite{NIPS2012_4824}) and games like GO is attributed to Deep Learning (DL) \cite{Goodfellow-et-al-2016} or Deep Neural Network (DNN). Restricted Boltzmann Machines (RBM) \cite{fisher2012ait} can be seen as stochastic neural networks that learn the input probability distribution in supervised as well as unsupervised manner and hence can be a powerful tools in detecting anomalies. In the context of digital twin, temporal RBM can be used to model multivariate timeseries, convolutional RBM is used to understand structure in timeseries, mean co-variance RBM to understand the covariance structure of the data and for dimensionality reduction \cite{Hinton504}. Shallow Autoencoders which are neural networks whose input and output layers are exactly the same and sandwich a layer with comparatively much reduced number of nodes have been in use for efficient linear data compression (similar to Principal Component Analysis (PCA)). However, just by stacking many more layers result in Deep Autoencoder. These have been used for very diverse applications like diabetes detection \cite{KANNADASAN2018}, action detection for surveillance \cite{ULLAH2019386}, feature learning for process pattern recognition \cite{YU20191}, denoising for speech enhancement \cite{LIU2018106}, fault diagnosis \cite{SHAO2017187}, social image understanding \cite{LIU2019}, low light image enhancement \cite{LORE2017650}. Convolutional Neural Network (CNN) is another DL method which although has achieved unprecedented success mainly in image classification (ImageNet, AlexNet, VGG, YOLO, ResNet, DenseNet etc.) exceeding human level accuracy, have also been extensively used in textual analysis. More recently Generative Adversarial Networks (GANs) \cite{NIPS2014_5423} where two networks called generators and discriminators are trained to outperform each other resulting in generators which can create data which are indistinguishable from real ones has achieved huge success. The method has huge potential in improving the data quality like upscaling image resolution \cite{LedigTHCATTWS16}, denoising \cite{Tripathi2018}, filling missing data \cite{Yoon2018gain} all of which are relevant in the context of digital twins. As a powerful Recurrent Neural Network (RNN), Long Short Term Memory (LSTM) network has demonstrated phenomenal success in modeling data involving time. Karim et al.~\cite{KARIM2019237} demonstrated the use of LSTM in timeseries classification. In \cite{SAGHEER2019203} LSTM was used to forecast petroleum production. By combining CNN and LSTM, Kim et al. in \cite{KIM201972} demonstrated its value in making predictions of the energy consumption in residential areas. The predictive capability of LSTM in particular will be instrumental in creating future scenarios in the context of digital siblings when direct observation data will be unavailable. The supervised algorithm in a digital twin context will be used for making predictions / projections or for conducting sensitivity / what-if analysis.

One of the shortfalls of supervised algorithms is the need of dependent variables (labeled data) which might not always be available as in the case of anomaly. Unbalanced or skewed data rarely results in reliable prediction models. Anomaly detection which is extremely important for online health monitoring of any system, the labeled data is rarely available in abundance to train a supervised algorithms. In such a situation unsupervised algorithms like self organized maps (SOM) \cite{KALTEH2008835} and clustering analysis \cite{Xu2015} (k-mean, t-SNE \cite{vanDerMaaten2008}) have better utility. Another important application of unsupervised algorithms like PCA and Deep Autoencoder can be for on-the-fly data compression for real-time processing, communication and control of the system under operation.

While supervised and unsupervised learning ML algorithms have been the most commonly employed algorithms in real applications, they are not of much use in the absence of enough data. Reinforcement Learning \cite{sutton1998introduction}, though in its infancy has the potential to aid in such a data-deprived situation. Reinforcement learning is an area of ML in which instead of learning directly from the data, an agent figures out an optimal policy to maximize its long term cumulative gain. It can be a key enabler for smart decision making systems in the technologies related to the Industry 4.0 (see also \cite{frank2006anatomy} for a nice discussion of the anatomy of a decision). For example, in \cite{rocchetta2019reinforcement}, the authors illustrated how a reinforcement learning framework might support operation and maintenance of power grid applications. The algorithms \cite{Silver2017mastering} without any prior knowledge and starting to play random gains, in 24 hours, achieved super-human level performance in chess, shogi as well as Go and convincingly defeated a work-champion program in each case. Since then the algorithm has been applied to solve more engineering problems like advanced planning of autonomous vehicles \cite{YOU20191},  lung cancer detection in medical treatment \cite{LIU20191}, smart agriculture \cite{BU2019500}, UAV cluster task scheduling \cite{YANG2019140}, chatbots \cite{CUAYAHUITL2019}, autonomous building energy assessment \cite{MASON2019300}.

Some advantages of the data-driven models is that they keep on improving as more and more data (experiences) are fed into them. The training part of the data-driven modeling might experience issues associated with instabilities. However, once trained the models are stable for making predictions. At the same time the data-driven models have some downsides too. Most complicated models of this class specially the ones based on DNN are uninterpretable. In safety critical applications like automatic drug injection in humans, guidance and navigation of autonomous vehicles or oil well drilling a black-box approach will be unacceptable. In fact the vulnerability of DNN have been been exposed beyond doubt in several recent works \cite{yuan2017AEA}, \cite{akhtar2018TAA}, \cite{xu2019adversarial}. These models can also be extremely biased depending upon the data they were trained on.

\subsection{Big Data Cybernetics: the art of steering}
\begin{figure}[htbp]
	\centering
	\includegraphics[width=\linewidth]{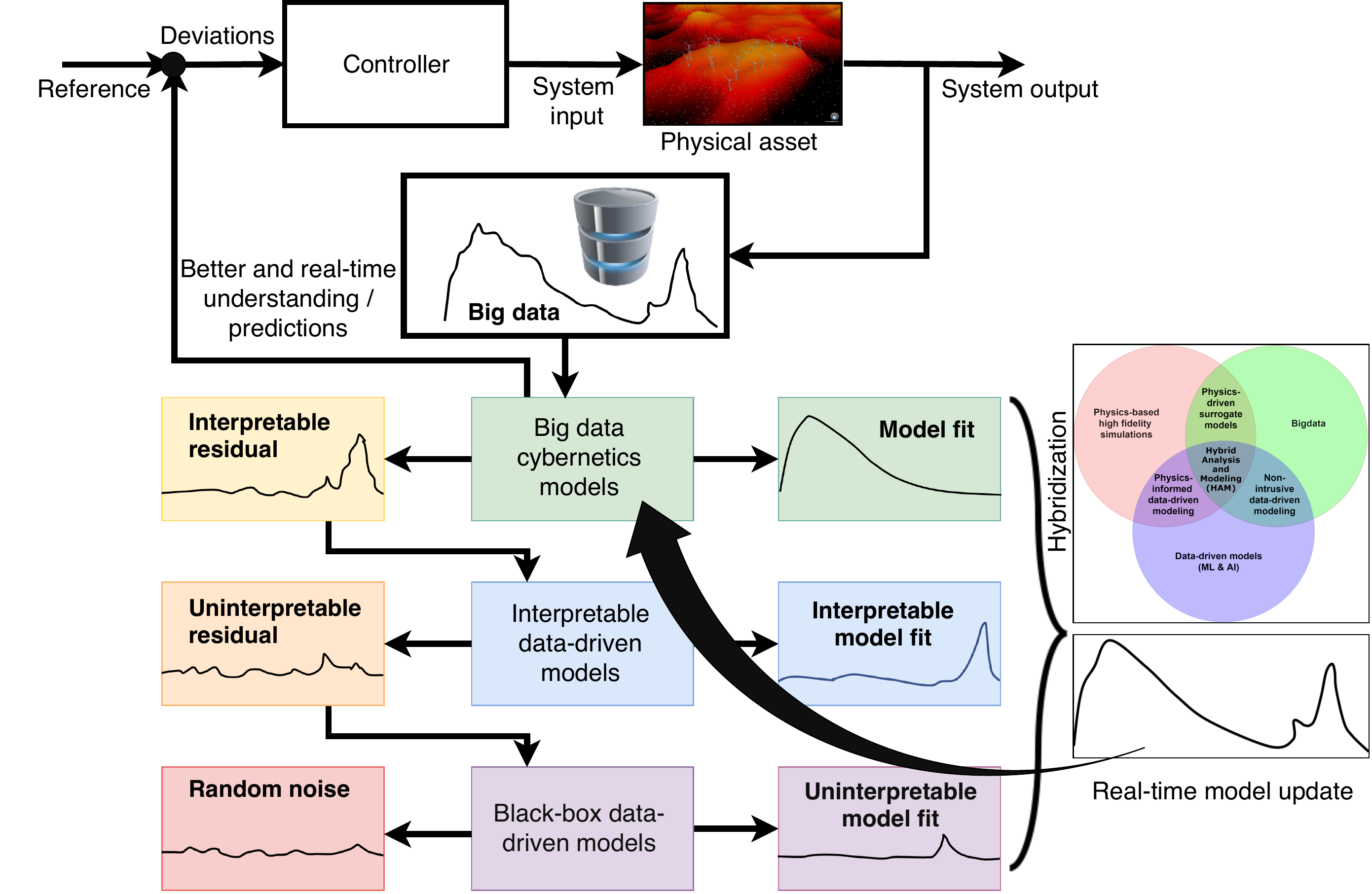}
	\caption{Big Data Cybernetics: The model starts as a physics based model that is based on first principles but as time progresses the model is continuously updated using knowledge generated from data}
	\label{fig:bdc}
\end{figure}
\begin{table*}
	\caption{\label{tab:PMvsDDM} Physics based modeling vs data-driven modeling}
	\begin{tabular}{p{8.5cm} p{8.5cm}}
		\hline
		Physics based modeling & Data-driven modeling\\ \hline
		$+$ Solid foundation based on physics and reasoning &$-$ So far most of the advanced algorithms work like black-boxes \\
		$-$ Difficult to assimilate very long term historical data into the computational models &$+$  Takes into account long term historical data and experiences \\
		$-$ Sensitive and susceptible to numerical instability due to a range of reasons (boundary conditions, initial conditions, uncertainties in the input parameters) &$+$  Once the model is trained, it is very stable for making predictions / inferences \\
		$+$ Errors / uncertainties can be bounded and estimated &$-$  Not possible to bound errors / uncertainties \\
		$+$ Less susceptible to bias & $-$  Bias in data is reflected in the model prediction \\
		$+$ Generalizes well to new problems with similar physics & $-$  Poor generalization on unseen problems\\
		\hline
	\end{tabular}
\end{table*}
Norbert Wiener defined cybernetics in 1948 as "the scientific study of control and communication in the animal and the machine," which can be easily linked to digital twins. The objective of cybernetics is to steer a system towards a reference point. To achieve this, the output of the system is continuously monitored and compared against a reference point. The difference, called the error signal is then applied as feedback to the controller which in turn generates a system input that can direct the system towards the reference point. At times the quantity of interest that is required to compare against the reference can not be measured directly and hence has to be inferred from other quantities that are easier and cheaper to measure. With an increase in computational power and availability of big data, there are two major improvement possible within the cybernetics framework. Firstly, the controllers can be improved by increasing its complexity to incorporate more physics. Secondly, the big data can be utilized for a better estimation of the quantity of interest. In order to address these two issues a new field of \textbf{big data cybernetics} shown in Fig.~\ref{fig:bdc} is proposed. It is an adaptation of the concept first conceived in \cite{martensbigdatacyb} at the Norwegian Univeristy of Science and Technology. In the figure, the first step is partial interpretation of big data using well understood physics-based models. The uninterpretable observation at this stage is termed interpretable residual and in a second step is modeled using an explainable data-driven approach. After the second step, again an uninterpretable residual remains which is modeled using more complex and black-box models preferably with some inbuilt sanity check mechanism. The remaining residual is generally noise which can be discarded. The three steps result in a better understanding of the data and hence improved models, provided, new approach can be developed to combine physics based modeling and data-driven modeling with big data. The steps are continuously looped with the availability of new streams of data (i.e., see Fig.~\ref{fig:bdc}). 
\begin{figure}[h]
	\centering
	\includegraphics[width=\linewidth]{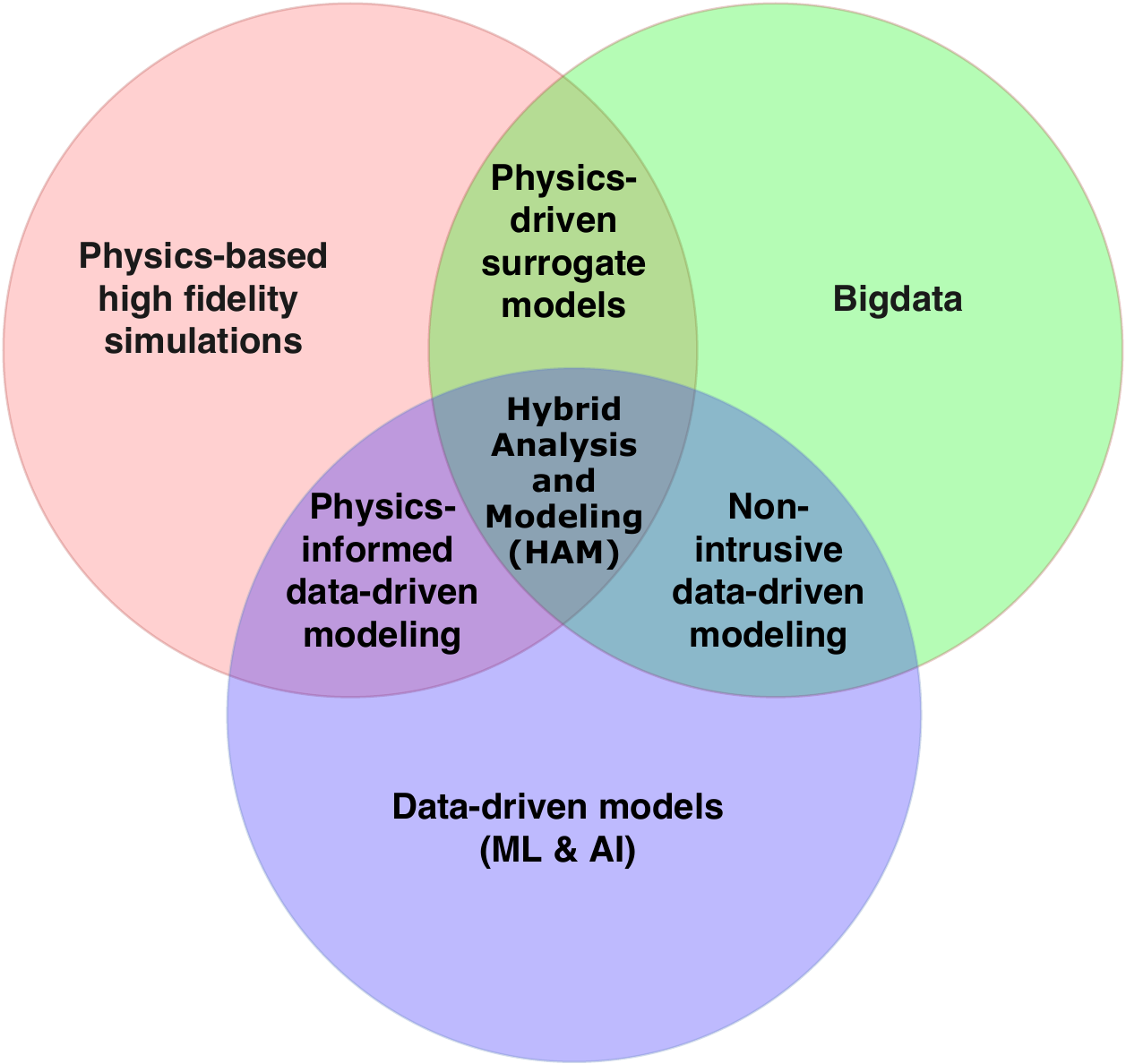}
	\caption{Hybrid Analysis and Modeling: Hybridization happens at the intersection of the three cricles}
	\label{fig:hm}
\end{figure}
The new approach should be intended towards removing the shortfalls of pure physics-based or pure data driven modeling approach (see Table~\ref{tab:PMvsDDM} for a summary). We call this new approach Hybrid Analysis and Modeling (HAM) and define it as a modeling approach that combines the interpretability, robust foundation and understanding of a physics based modeling approach with the accuracy, efficiency, and automatic pattern-identification capabilities of advanced data-driven ML and AI algorithms. It is possible to place HAM approaches at the intersections of big data, physics based modeling and data driven modeling as shown in Fig.~\ref{fig:hm}.   

\subsubsection{Data Assimilation}
Data assimilation (represented by the intersection of big data and physics-based high fidelity simulations in Fig.~\ref{fig:hm}) is a well-established mathematical discipline that combines computational models with observations. It is a geoscience terminology and refers to the estimation of the state of a physical system given a model and measurements. In other words, it is the process of fitting models to data. In engineering fields the terms \emph{filtering}, \emph{estimation}, \emph{smoothing}, \emph{prediction} are often used \cite{moheimani1998robust,pitt1999filtering,arulampalam2002tutorial,doucet2009tutorial}. 	
In practice, many techniques in data assimilation can be derived both from a deterministic and stochastic point of view. There are many presentations of the topic in books, review and tutorial articles (e.g., we refer to \cite{ghil1991data,bennett2005inverse,lewis2006dynamic,evensen2009data,majda2012filtering,sarkka2013bayesian,park2013data,law2015data,van2015nonlinear,lakshmivarahan2017forecast,fearnhead2018particle} for detailed mathematical descriptions). Although data assimilation approaches initially gained popularity in numerical weather forecasting \cite{malanotte1996oceanographic,talagrand1997assimilation,bertino2003sequential,tsuyuki2007recent,helmert2018review,navon2009data,crestani2013ensemble,houtekamer2016review}, they are now routinely used in fields as diverse as finance, medical physics, signal processing, computer vision, robotics and navigation \cite{tanizaki1994prediction,duan1994maximum,bouchouev1997inverse,bouchouev1999uniqueness,wells2013kalman,javaheri2003filtering,rachev2008bayesian,sarvas1987basic,schmidt1999bayesian,soleimani2008computational,ansley1990filtering,kam1997sensor,okuma2004boosted,buch2011review,sarkka2013gaussian,paull2013auv,liu2015real,liu2018visual}, and these techniques offer a great promise in many digital twin applications. 

In an operational atmospheric forecast, data assimilation deals with very large-scale state space systems and typically enables initialization of models by statistically combining information from short-range model forecasts and observations, based on the estimated uncertainty of each \cite{kalnay2003atmospheric}. Data assimilation methods differ primarily in their treatment of the model background error and the methods for solving the analysis equations \cite{hacker2018challenges}. Both variational (adjoint methods) and sequential methods (Kalman filters) have been successfully used in operational weather centers to minimize the error between forecasting trajectory and noisy observation data. Dedicated comparisons of four-dimensional variational data assimilation (4D-VAR) and ensemble Kalman filter (EnKF) approaches can be found in \cite{lorenc2003potential,kalnay20074,fairbairn2014comparison}. Their variants as well as hybridized approaches have been also introduced in atmospheric science and numerical weather predictions \cite{lorenc2015comparison,tian2019adjoint}. We refer to recent studies on the use of the Kalman filter \cite{brandtstaedter2018digital} and particle filter \cite{li2017dynamic} in the digital twin applications.    


\subsubsection{Reduced Order Modeling}
There are a great number of high-dimensional problems in the field of science (like atmospheric flows) that can be efficiently modeled based on embedded low-dimensional structures or reduced order models (ROMs). This family of model can be best described at the intersection of physics-based high fidelity simulations and data-drive models as shown in the Fig.~\ref{fig:hm}. These ROMs often trade a level of accuracy for much greater computational speed. This ROM concept (also called surrogate modeling or metamodeling) allows one to emulate complex full-order models or processes, and lies at the interface of data and domain specific sciences. It has been systematically explored for decades in many different areas including computational mechanics \cite{meyer2003efficient,quarteroni2011certified,kerfriden2011bridging,xia2014reduced}, combustion \cite{goussis2011model,anand2011surrogate,hemchandra2012premixed}, cosmology \cite{tegmark1997karhunen,purrer2014frequency,field2014fast}, electrodynamics \cite{wittig2006model,koziel2012model,hochman2014reduced,kersaudy2015a}, meteorology \cite{maasch1990a,selten1995an,crommelin2004strategies,franzke2005low}, fluid mechanics \cite{sirovich1987turbulence,aubry1988dynamics,willcox2002balanced,noack2003hierarchy,rowley2005model,mezic2005spectral,loiseau2018sparse}, heat transfer \cite{yesilyurt1995surrogates,park1996use,tallet2016optimal} and various other multiphysics processes \cite{osio1996an,qian2006building,amsallem2008interpolation,neron2010proper,boulakia2012reduced,kerfriden2013partitioned,ghavamian2017pod}, as well as systems and control theories \cite{moore1981principal,juang1985eigensystem,kirby1990application,everson1995karhunen,chaturantabut2010nonlinear,peitz2019koopman} in order to provide computational feasible surrogate models. Although the concept can be traced back to the works done by Fourier (1768-1830), there exist many recent monographs \cite{schilders2008model,buljak2011inverse,noack2011reduced,holmes2012turbulence,koziel2013surrogate,quarteroni2014reduced,hesthaven2016certified,kutz2016dynamic} and review articles \cite{berkooz1993proper,okino1998simplification,lucia2004reduced,queipo2005surrogate,hannachi2007empirical,chinesta2011short,razavi2012review,brunton2015closed,benner2015survey,asher2015a,rowley2017model,taira2017modal,bhosekar2018advances,lu2019review}. In practice, ROMs (i.e., emulators) have great promise for applications especially when multiple forward full-order numerical simulations are required (e.g., data assimilation \cite{houtekamer1998data,houtekamer2001sequential,daescu2007efficiency,cao2007reduced,he2011use,law2012evaluating}, parameter identification \cite{martin2005use,xun2013parameter,kramer2017feedback,fu2018pod}, uncertainty quantification \cite{mathelin2005stochastic,najm2009uncertainty,galbally2010non,biegler2011large,smith2013uncertainty,renardy2018parameter,moore2018predictive}, optimization and control \cite{graham1999optimala,graham1999optimalb,ravindran2000reduced,ito2001reduced,mcnamara2004fluid,jeong2005efficient,bergmann2008optimal,loiseau2019pod}). These models can quickly capture the essential features of the phenomena taking place, and we often might not need to calculate all full order modeling details to meet real-time constraints.  Another advantage can be realized for developing surrogate models stationary parameterized systems, where full order models typically require many pseudo-iterations to converge to a steady-state at each control parameter \cite{ly2001modeling,ding2008fast,samadiani2010reduced}. Therefore, reduced order modeling has been considered as a key enabler to compress high fidelity models into much lower dimensions to alleviate heavy computational demand in digital twin technologies \cite{hartmann2018model}. ROM enables reusing simulation models from the early phases of product development in later product lifetime phases, especially during the product operation phase \cite{keiper2018reduced}. 

Most of the conventional ROMs require the exact form of the underlying partial differential equations (PDEs) involved to explain the physics phenomena (hence called intrusive). This is sometimes impractical due to Intellectual Property Rights or issues related to an incomplete understanding of the underlying processes. Alternatively, nonintrusive data-driven models (lying at the intersection of big data and data-driven models in Fig.~\ref{fig:hm}) have emerged recently with the democratization of computational power, explosion of archival data, and advances in ML algorithms \cite{audouze2013nonintrusive,mignolet2013review,xiao2015non,peherstorfer2016data,hesthaven2018non,hampton2018practical,chen2018greedy,xiao2019domain,wang2019non,san2019an,rahman2019a}. These approaches mostly involve matrix operations, which can be very efficiently parallelized on affordable GPU and TPU giving several orders of magnitude speedup as required in real-time applications. One of the main advantages of a nonintrusive approach is its portability, which results from the fact that it does not necessarily require the exact form of the equations and the operators or methods used to solve them. This makes the approach applicable to experimental data where the equations are often not well established or have huge uncertainties involved in their parameters. Together with their modularity and simplicity, nonintrusive models offer a unique advantage in multidisciplinary collaborative environments. It is often necessary to share the data or the model without revealing the proprietary or sensitive information. Different departments or subcontractors can easily exchange data (with standardized I/O) or executables, securing their intangible assets and intellectual property rights. Furthermore, nonintrusive approaches are particularly useful when the detailed governing equations of the problem are unknown. This modeling approach can benefit from the enormous amount of data collected from experiments, sensor measurements, and large-scale simulations to build a robust and accurate ROM technology.
Once ROMs are constructed its operational execution cost is usually negligible in PDE constrained optimization problems. Besides, another advantage can be viewed in such a way that the optimization task may be separated from the generation of data for a ROM via a design of experiments (DOE) approach or other interpolation approaches \cite{xiao2010model,simpson2001metamodels}. The statistical approach proposed by Krige \cite{krige1951statistical} has become highly popular in such multidimensional interpolation problems (known as Kriging or Gaussian Process Regression \cite{williams2006gaussian}). Of course, we have to take into account the cost of creating an initial database requiring large-scale high-fidelity simulations within complex domains discretized by finite volumes or finite elements. The authors, in their recent studies \cite{fonn2017astep,fonn2019fastdiv,siddiqui2019finitevolume}, make ROM inter-operatable between different discretization techniques. In addition, the use of radial basis functions (RBF) in finite elements can be found in \cite{noor1980reduced} for nonlinear structural analysis. A description of radial basis approximation in PDE systems can be found in \cite{rozza2007reduced}. A hybrid reduced basis and finite element method has been proposed as a component-to-system model order reduction approach, which is considered as one of the key enablers in digital twin technologies \cite{ballani2018component}. An application of RBF to construct a noninstrusive ROM for data assimilation applications can be found in \cite{xiao2018parameterised}.

\subsubsection{Hardware and Software in the Loop}

Hardware-in-the-loop (HIL) simulation approach, a well-established concept of using a physical object during system development phase in control engineering \cite{isermann1998hardware,bacic2005on,gietelink2006development}, can be also considered as a promising approach towards near real-time predictions within digital twin platforms (e.g., see also software-in-the-loop (SIL) simulation studies for early design processes \cite{kwon1999real,demers2007generic,jeong2016software}). For example, a real-time co-simulation concept for HIL simulations has been presented in \cite{scheifele2019real} considering the virtual commissioning of production systems.

\subsubsection{Other Hybridization Techniques}
As the constructive criticism of Albert Einstein and many others had helped to develop a better understanding of quantum mechanics \cite{ballentine1972einstein,whitaker2006einstein,kumar2008quantum}, we believe the valuable criticisms coming from domain scientists might help to develop more interpretable ML methodologies for scientific applications. For example, we refer to recent surveys of ML in domain-specific areas of molecular and material science \cite{butler2018machine} and fluid mechanics \cite{brunton2019machine}.

Complete replacement of equations with ML algorithms for gaining computational efficiency when generalization is not a necessity (e.g., we refer to \cite{lee1990neural,lagaris1998artificial,lagaris2000neural,tompson2017accelerating,tang2017study,sirignano2018dgm,kim2019deep,wiewel2019latent} and references cited therein for speeding up numerical solvers for real-time applications) can be used in the context of digital twin when the same process is to be monitored time and again. However, this approach will fail in unexpected situation because the model only learns to interpolate and not extrapolate.

A better approach on the integration of ML to physical processes has been demonstrated intuitively in \cite{maulik2019subgrid} where the known physics is modeled using the established governing equations while the unknown physics is modeled using black-box DNN or LSTM networks. When combined, the black-box part improves the overall accuracy of the combined model and the equation based part puts an inbuilt sanity check mechanism to detect unexpected behavior of the black-box part. In their perspective \cite{reichstein2019deep}, the authors, considering an Earth system model, charted a comprehensive integration scheme that can be done through (i) improving parameterization, (ii) replacing physical submodel with machine learning, (iii) analysis of model-observation mismatch, (iv) constraining submodels, and (v) developing surrogates. It has been concluded that a joint modeling approach offers viable solutions to tackle challenges such as interpretability, physical consistency, noise, limited data, and computational demand, when these theory-driven physical process models combined with the versatility of data-driven ML.      

Another approach to improving the interpretability of any ML algorithm is to reduce the complexity of the algorithm. One way of doing this is to engineer new features based on domain knowledge and informing the ML about it. Since, the algorithm is relieved of the additional task of rediscovering these features, one can work with simpler algorithms with far better accuracy. This approach was demonstrated in the context of material discovery in  \cite{Elemnet2018}, \cite{tabibthermoelectric2018}.

More work on hybrid analytics or hybrid physics-AI modeling can be found in \cite{krasnopolsky2005new,wan2018data,pathak2018hybrid,rahman2018hybrid,vlachas2018data}. For example, we refer to the works done in \cite{krasnopolsky2006complex,krasnopolsky2006new} using such hybrid approaches combining deterministic and ML components in general circulation models (GCMs) for applications to climate modeling and weather predictions. The processes in GCMs can be split into two parts referring to model dynamics (i.e., set of 3D equations of motion governed by PDEs for mass, momentum, energy etc) and model physics (i.e., source terms for parameterizing atmospheric radiation, turbulence, precipitation, clouds, chemistry, etc). The hybrid approach has a good motivation for emulating model physics, since this part often dominates the total model computation time. Similar gray-box approaches have been gaining popularity in various other fields too (e.g., hydrology \cite{nourani2014applications}, material forming \cite{ibanez2019hybrid}, bioprocess \cite{zhang2019hybrid}, built environment \cite{gray2018hybrid}, petroleum \cite{liu2019accelerated}, reactor design \cite{mosavi2019prediction} and quantum chemistry \cite{zhang2018potential}).

Another interesting way of combining physics based modelling and big data is the approach advocated by Soize and Farhat~\cite{Soize2017anp} and~\cite{Soize2019plf}. They exploits available data to adapt the subspace in which the solution of the problem formulated using the computational model is searched. The method is innovative as it discover some form of the information or knowledge encapsulated in data, instead of the common approach to adapt model parameters. The resulting nonparametric probalistic reduced order method enables a sound mathematical/statistical based combination of physics based model and data highly relevant in a digital twin context.

\subsubsection{Physics-informed ML}
When it comes to utilizing ML algorithms in cases where the underlying physical process of high dimension, some of the challenges include incorporating physical laws within the learning framework, producing solutions that are interpretable, addressing nonlinearities, conservation properties, and dealing with the massive amount of data needed for training. For example, in mathematical models governed by PDEs, the authors in \cite{raissi2019physics} demonstrated a different approach of programming physics directly into the DL algorithm. They did so by using the residual of the governing equations to regularize the cost function that can be optimized in any ML algorithm. In \cite{han2018solving}, the PDEs are reformulated as backward stochastic differential equations and the gradient of unknown solution is approximated using neural networks. This is inspired from the deep reinforcement learning with gradient acting as the policy function. Physics-informed ML ideas have been also utilized in various areas including inorganic scintillator discovery \cite{pilania2018physics}, fluid dynamics \cite{ling2016reynolds,wang2017physics,wu2018physics}, projection-based model reduction \cite{swischuk2019projection}, cardiovascular system modeling \cite{kissas2019machine}, and wind farm applications \cite{howland2019wind}. In \cite{ruden2019informed}, the authors have presented a comprehensive review on the taxonomy of explicit integration of knowledge into ML. As generously highlighted by Martens in his perspective paper  \cite{martens2015quantitative}, most domain-specific software tools can be improved by fresh impulses from the developments in computer and cognitive sciences, and in return, many specific disciplines have a lot to give to these sciences as well as to other fields. Therefore, the concept of Physics/Knowledge/Science-informed ML might offer a powerful transformation to combine practice and theory (e.g., please see \cite{faghmous2014theory,wagner2016theory,karpatne2017theory,jia2019physics,greve2019data} for promoting theory-guided ML in various disciplines). For example, a symmetry enforcing approach to learn salient features from a sparse data set has been introduced using a knowledge injection concept \cite{bergomi2019towards}, that might be a key enabler in developing digital twin technologies.



\subsubsection{Compressed Sensing and Symbolic Regression}   
Compressed sensing (CS) has been applied to signal processing in seeking the sparsest solution \cite{donoho2006compressed,candes2008introduction,needell2009cosamp}. Sparsity-promoting optimization techniques (e.g.,  least absolute shrinkage and selection operator (LASSO) and its generalizations) often perform feature selection through $L_1$ penalty added to recover sparse solutions \cite{tibshirani1996regression,james2013introduction,hastie2015statistical}. Ridge regression is another regularized variant where $L_2$ penalty is added to objective function \cite{hoerl1970ridge,myers1990classical,murphy2012machine}. Elastic nets combine the strengths of the LASSO and ridge approaches \cite{zou2005regularization,mol2009elastic,friedman2010regularization}. Sequentially thresholded ridge regression (STRidge) can be also used in more complex systems with correlated basis functions \cite{rudy2017data}. 

In recent years, the use of ML methods has complemented the need for formulating mathematical models, and allowed accurate estimation of observed dynamics by learning automatically from the given observations and building models. For example, sparse regression has been applied for recovering several nonlinear canonical PDEs \cite{rudy2017data}. These sparse optimization techniques have been used to recover the underlying basis from many candidate featuring terms. The computational time of these approaches are small and in general they can handle large datasets. One of the key concern with these techniques are that they perform poorly on data corrupted with noise. In addition, evolutionary approaches are proposed at another end of spectrum to leverage randomness as virtue in recovering models. Symbolic regression (SR) methods explore a function space, which is generally bounded by a preselected set of mathematical operators and operands, using a population of randomly generated candidate solutions. A seminal work by Schmidt and Lipson~\cite{schmidt2009distilling} has demonstrated that the SR approach can be used to determine the underlying structure of a nonlinear dynamical system from data. This approach is crucial where data is abundant (e.g., geophysical flows, finance, neuroscience) but the accurate physical law/model is not available. When it comes to large-scale datasets, gene expression programming (GEP), for example, has been applied to recover turbulence closures \cite{weatheritt2016novel}. The authors tailored original GEP algorithm \cite{ferreira2006gene} to handle tensor regression and recover complex functional forms. Recently, a mixed approach was proposed by Both et al.~\cite{both2019deepmod} where the feed forward neural networks and sparse regression techniques were combined to handle noisy data.  These tools can be exploited as a potential data-driven tool for recovering hidden physical structures or parameterizations representing high-dimensional systems from data in digital twin applications.

\subsection{Infrastructure and platforms}	

\subsubsection{Big data technologies}
\begin{figure}[htbp]
	\centering
	\includegraphics[width=\linewidth]{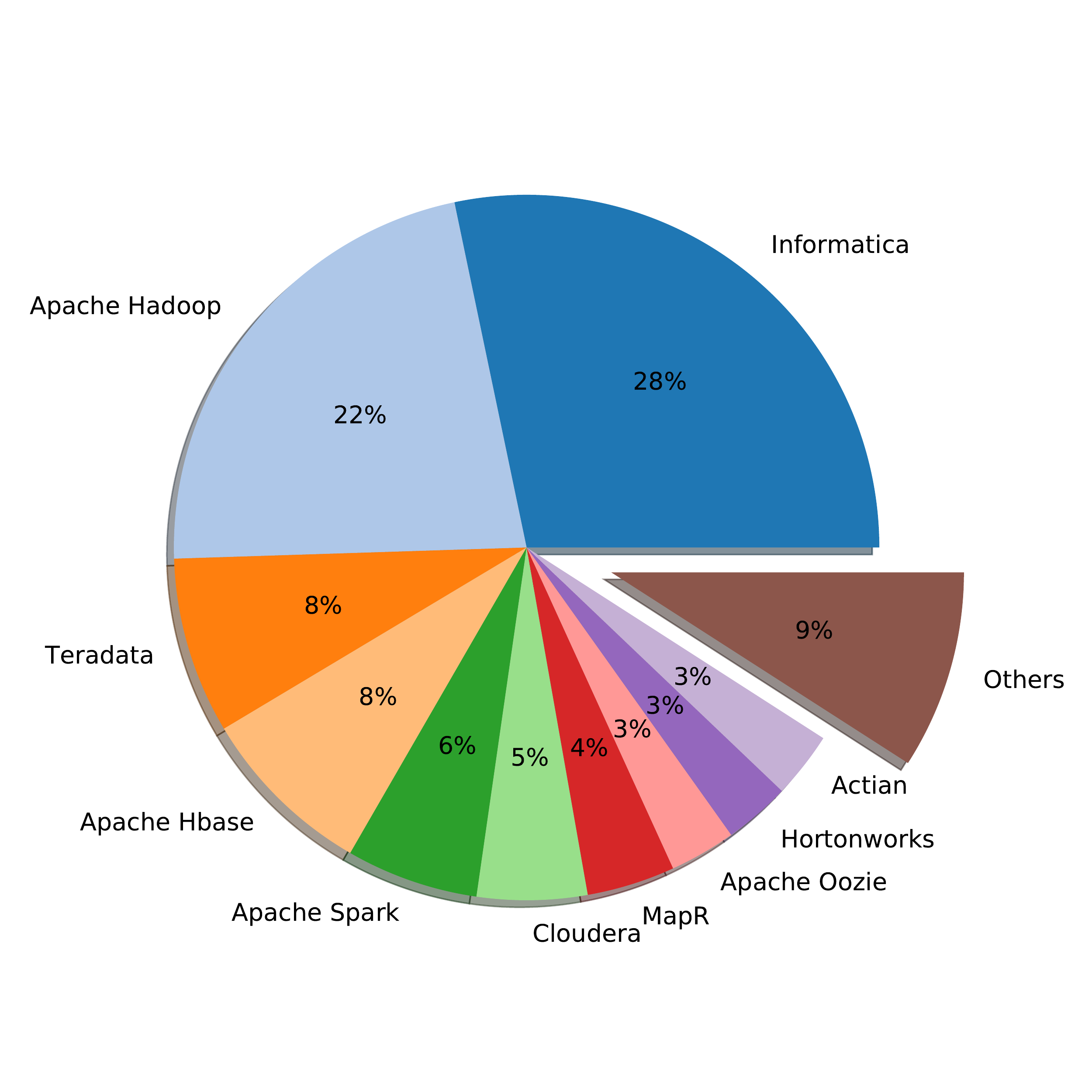}
	\caption{Market share of big data products}
	\label{fig:bigdata}
\end{figure}

The infrastructure for storing and processing high volume data has been advanced considerably over the last decades. Many available platforms are available to handle big data projects in terms of blending, integration, storage, centralized management, interactive analysis, visualization, accessibility and security. Many IT vendors benefit from Hadoop technology \cite{white2009hadoop,zikopoulos2011understanding,landset2015survey,mazumder2016big}, which allows us to execute tasks directly from its hosting place without copying to local memory. According to enlyft's report considering 136,929 companies \cite{enlyft}, market share statistics of big data products used in these companies can be summarized in Fig.~\ref{fig:bigdata}. Indeed, there are many survey articles to discuss various aspects of big data including not only state-of-the-art technology and platforms, but also algorithms, applications and challenges  \cite{sagiroglu2013big,chen2014big,chen2014data,zhang2015memory,wang2015learning,singh2015survey,tsai2015big,hashem2015rise,jiang2016energy,qiu2016survey,zhou2017machine,skourletopoulos2017big,sangeetha2017survey,oussous2018big,zhang2018survey}. More recently, Kaufmann~\cite{kaufmann2019big} introduced a big data management concept to address the critical needs of digital twin applications.  

\subsubsection{IoT Technologies}


The IoT is becoming increasingly popular to develop smart technologies in several sectors ranging from healthcare to agriculture, from transportation to energy. The availability of high-speed Wi-Fi internet connection, digital machines, low-cost sensors along with the development of ML algorithms to perform real-time analysis has contributed to enabling IoT \cite{madakam2015internet,vcolakovic2018internet}. The leading IoT vendors provide reliable and innovative platforms to set up IoT devices for various applications \cite{ray2016survey}. 

Amazon Web Services (AWS) offers various IoT solutions \cite{murty2008programming}. For example, AWS IoT finds its application in biomedical research to identify biological insights which were not known before \cite{fusaro2011biomedical}. The high volume of complex biomedical data is being generated using the state-of-the-art high-throughput machines and this data can be processed and stored efficiently using cloud computing.  In their work \cite{fusaro2011biomedical}, the cloud computing framework was applied to genome mapping which uses a large amount of next-generation sequencing (NGS) data. This task can be easily distributed over a computer cluster and cloud computing is an ideal platform for this, as it does not have limitations of high-performance computing (HPC) systems such as job queues and unsupported software. Another application of IoT is in sensor network technologies which are used extensively in environmental monitoring and industrial automation. Cloud computing often provides a flexible computational model that is perfectly suitable for unpredictable demands generated by environmental sensor networks. For example, in \cite{lee2010extending}, Amazon's elastic compute cloud (EC2) was applied for processing dynamic data collected by sensor networks in environmental applications.

IoT has also been used in developing smart home technologies in which the operating conditions such as humidity, temperature, luminosity, etc are controlled with minimum user's intervention. In \cite{soliman2013smart}, IoT concepts and cloud computing were integrated to facilitate smart home implementations. Their architecture uses microcontroller-enabled sensors to monitor home conditions, actuators for performing required actions, and Google App Engine platform for creating and setting up their web application. In \cite{hussain2011mining}, the authors used Twitter application to collect data, and presented Google App Engine as a possible solution for mining and analyzing the large amount of data.

Microsoft launched Azure IoT Suite, a platform that enables end-users to communicate with their IoT services and devices, exchange data, and process it in a convenient way \cite{ammar2018internet}. Many other specified platforms have been built on top of Azure for different applications. Forsstr{\"o}m and Jennehag \cite{forsstrom2017performance} investigated a system built by combining Azure IoT hub and the open plant communication universal architecture (OPCUA) for a realistic industrial situation tracking an industrial component with 1500 sensors. For agricultural purposes, Vasisht et al.~\cite{vasisht2017farmbeats} presented FarmBeats, an end-to-end IoT platform for data-driven agriculture which uses Microsoft Azure IoT Suite for cloud services and passing data summaries to storage. FarmBeats is capable of handling data from various sensor types with different bandwidths. Also, it is sufficiently reliable even in cases of power and internet outages from bad weather. For healthcare applications, Manashty et al.~\cite{manashty2015healthcare} proposed a healthcare event aggregation lab (HEAL) model in order to detect health anomalies as accurately and fast as possible. A prototype of HEAL was implemented and tested on Azure. Das et al.~\cite{das2015application} proposed a cloud based approach to make users aware of probable health risks of the flickering of surrounding artificial light. A small-length video is recorded using phone's camera. It is then uploaded to Azure cloud storage, where it can be further processed and analyzed to give informed instructions.

SAP cloud platform helps to implement IoT applications. Koduru et al.~\cite{koduru2019smart} proposed a framework based on SAP cloud for implementing a smart irrigation system taking water supply from borewells and weather conditions into account. Tsokov and Petrova-Antonova~\cite{tsokov2017ecologic} proposed EcoLogic, a solution for real-time monitoring of running vehicles and their respective emissions' levels, in order to keep the carbon emissions within acceptable limits through smart notifications and management of vehicle’s power. The system combines hardware module that collects data related to vehicles' emissions with cloud-based applications for data processing, and analysis using SAP cloud platform. 

Salesforce IoT which is based on software delivery paradigm where users access the software through personal web browsers and vendors host the business software. This software delivery paradigm is categorized under software-as-a-service (SaaS) built on cloud computing \cite{peng2018boat}. This paradigm helps in scalability, efficient maintenance, efficient management and lower resource costs. Salesforce IoT platform is designed to take in the massive amount of data generated by devices, sensors, websites and applications and initiate actions for real-time responses. Some real-time applications might include wind turbines that automatically adjusts its parameters based on live weather data, passengers whose connecting flights are delayed could be rebooked before they have landed \cite{LI2018,serrano2013resource,lucero2016iot}. Salesforce customer relationship management (CRM) software package was used by  organizations to anticipate customer needs based on their past requirements and choices \cite{manchar2017salesforce}. Such platforms have been also deployed in governmental organizations \cite{onyegbula2011understanding}. 

Oracle IoT is a cloud-based service for business implementations and uses software delivery paradigm called platform-as-a-service (PaaS) \cite{kuila2019analytical,achary2017internet}. Among many applications, Oracle IoT has been used for analyzing data regarding user's venue recommendations such as user’s current geographical location, weather and traffic conditions, cost, service or distance of the venues, as well as social network user reviews of venues \cite{Margaris2017}.

Cisco is one of the early business groups to realize the birth and potential of IoT based platforms \cite{evans2012internet,singh2015internet,darshan2015comprehensive}. Okafor et al.~\cite{okafor2017leveraging} used Cisco Nexus platform to develop a scalable IoT datacenter management (cost-effective processing, storing, and analysis of large volumes of data) for fog computing. In \cite{chiang2016fog}, the authors summarize the opportunities and challenges of fog computing in the networking context of IoT. A fog assisted IoT system has been considered as a health monitoring system in smart homes \cite{verma2018fog}. In their recent work, Bakhshi et al.~\cite{bakhshi2018industrial} discussed industrial security threats and concerns for both Cisco and Microsoft Azure IoT architectural models (for data accumulation and abstraction layers of those model architectures) and then suggested some security considerations. In fact, the security and privacy concerns are the major challenges of implementing any IoT devices \cite{o2016insecurity}.

Bosch IoT suite provides not only services through cloud connections but also the on-premise services with no third party access to the local data \cite{schmid2016architecture,swetina2014toward}. In \cite{broring2017enabling}, the authors introduced an architectural model to successfully establish such IoT ecosystem through platform interoperability. An example of integrating IoT devices provided by Bosch IoT suite into a business process has been discussed in details in \cite{friedow2018integrating}.

Using the IBM BlueMix cloud computing platform, an IoT technology has been applied to build a smart car system that monitors and the car's activity and informs the owner or technician about the routine maintenance to ensure safe and efficient driving \cite{husni2016applied}. 
In \cite{kumar2017air},  real-time air quality parameters such as humidity, temperature, and particulate matter were measured using different sensors. Data storage, data management, and data analysis were performed using the IBM Bluemix cloud to inform about the necessary decisions promptly.

\subsubsection{Communication Technologies}
A reliable working of any digital twin will require information arising from different components to reach its intended target on time. For example, during a robotic surgery, the action of a surgeon in a digital operation theater should manifest into action in reality without any latency. With so many sensors comes the problem of fast data communication. The state-of-the-art communication technology like 4G will run into problems as more and more devices will start sharing the limited radio-frequency spectrum. 5G technology with a wider frequency range might accommodate many more devices but that requires communication at much higher frequencies (30-300GHz) compared to the currently used frequencies by our mobile network. Such high frequencies waves called Millimeter Waves \cite{han2019millimeter} can not penetrate obstacles readily and hence more and more miniaturized base stations called Small Cells operating on low power will be placed spanning the whole relevant area. These base stations can also support hundreds of ports for many more antennas. The technology is called Massive MIMO. The large number of antennas mean more signal interference which will be corrected using the Beamforming technology \cite{han2019millimeter} which is a traffic-signaling system for the cellular base station which optimizes the data-delivery route to a particular user. Finally, for multiple communication at the frequency full duplexing is being looked into. The 5G technology will form the backbone of any digital twin owing to their ultralow latency and unprecedented speed and is already being used in the world's first 5G cinema in Oslo. 

Long Range Wide Area Network (LoRaWAN) technology has been gaining a great attention for IoT applications, while keeping network structures and management simple \cite{nolan2016evaluation,adelantado2017understanding,kwasme2019rssi}. Unmanned Aerial  Systems (UAS) cooperations have also recently attracted considerable attention to assist ground base stations in the case of crowded public events, natural disasters, or other temporal emergencies that require additional needs for network resources \cite{chetlur2017downlink,turgut2018downlink,nolan2018coordinated}. For example, UAS based sensing systems have demonstrated value for flexible in-situ atmospheric boundary layer observations to improve weather forecasting \cite{elston2011evaluation,bonin2013observations,jacob2018considerations}. 

\subsubsection{Computational Infrastructures} 
According to Moore's Law, the performance and functionality of computers/processors can be expected to double every two years due to the advances in digital electronics. Therefore,	in scientific computing applications, the speeds of both computation and communication have substantially increased over the years. However, the communication becomes more challenging as the number of processing elements increases and/or the number of grid points decreases within a processing elements, which constitutes a major bottleneck at upcoming exascale computing systems \cite{dongarra2011international,reed2015exascale,donzis2014asynchronous}. Unless we face up to a new transistor technology (e.g., see \cite{hills2019modern}) to replace current metal-oxide semiconductor transistor technologies, this doubling trend saturates as chip manufacturing sector reaches the limits of the atomic scale. This leads to more effective use of transistors through more efficient architectures (e.g., see \cite{vetter2019extreme} for a recent discussion in extreme heterogeneity). The HPC community has started to move forward to incorporate GPU based accelerators and beyond (e.g., TPUs \cite{jouppi2017datacenter}) for not only graphics rendering but also scientific computing applications. This heterogeneity shift becomes even more crucial in future since there is a rapid increase in the usage of high-productivity programming languages (e.g., Matlab, R, Python, Julia) among engineers and scientists. In a recent article \cite{katz2018community}, the authors reviewed synergistic activities among major software development organizations considering challenges such as productivity, quality, reproducibility, and sustainability.

As an enabling technology for digital twins, cloud computing utilizes computing power at remote servers hosted on the Internet to store, manage, and process information/data, rather than a local server or a personal computer \cite{buyya2009cloud,zhang2010cloud,pallis2010cloud}. Edge computing utilizes computing power at the edge/nodes of the networks, to completely or partly to store, manage, and process	information/data locally \cite{shi2016promise,shi2016edge,mao2017survey,mach2017mobile}. Fog computing is a blend of cloud and edge computing where it is unknown to the user where in the network the data is stored, managed and	processed, and the load is distributed automatically between remote servers and local resources \cite{dastjerdi2016fog,dastjerdi2016fog}. In \cite{gupta2017ifogsim}, the authors developed a simulator to model IoT using fog environments and measured the impact of resource management techniques in latency, network congestion, energy consumption, and cost. We refer to \cite{al2015energy} for the energy management and scalability of IoT systems using fog computing. In \cite{aazam2018offloading}, the authors reviewed offloading in fog computing in IoT applications. Furthermore, granular computing paradigm \cite{skowron2016interactive} offers a general framework based on granular elements such as classes, clusters, and intervals and might be quite useful in big data applications \cite{zhang2012rough,huang2017dynamic,wang2017an}. It is also anticipated that emerging quantum computing systems will offer phenomenal capabilities for solving data-intensive complex problems (e.g., computational sciences, communication networks, artificial intelligence \cite{peters2019quantum}). 

\subsubsection{Digital Twin Platforms} 

Kongsberg Digital is a company delivering PaaS and SaaS services for energy, oil and gas, and maritime industries. Kongsberg is partner with some key cloud vendors, including Microsoft Azure, to provide enabling information technology services \cite{ullevik2017impact}. In 2017, Kongsberg launched its open digital ecosystem platform, called KognifAI \cite{Kognifai}. It combines a group of applications on the cloud focusing on optimal data accessibility and processing. KognifAI is built on cybersecurity,  identity, encryption, and data integrity \cite{Kongsberg_Hart}. Moreover, its main infrastructure can be utilized to easily scale applications and services \cite{Kongsberg_OceanNews}. KognifAI offers digital twin solutions in maritime, drilling and wells, renewable energies, etc \cite{Kognifai_sol}. Kongsberg dynamic digital twin combines safety with fast prototyping and implementation, and connects offshore and onshore users in oil and gas industry \cite{Kongsberg_Energy}. It can provide a model not only to represent an existing physical system, but also a planned installation (e.g., greenfield digital twin \cite{Kognifai_greenfield}) or maintenance and repair (e.g, brownfield \cite{Kognifai_brownfield}).

In 2018, MapleSim, a product of MapleSoft, added new features for developing digital twins \cite{Mapple}. MapleSim is powerful in creating an accurate dynamic model of the machine based on CAD data, with all forces and torques included. The major focus of MapleSim digital twin module is to implement virtual plant models that do not necessarily require expert knowledge. To test motor sizing in a new screwing machine at Stoppil Industrie, a digital twin of the machine was created using MapleSim \cite{Mapple_screwing}. The initial motor size was found to be undersized by a factor of 10, which would have caused machine failure and excessive losses. MapleSim was integrated into B\&R automation studio to facilitate the use of digital twin for machine development \cite{Mapple_BR}.  MapleSim in conjunction with B\&R software were used to build model-based feedback during motor sizing for an injection molding machine \cite{Mapple_injection}.

Cognite provides the full-scale digital transformation services to heavy industries such as oil and gas, power \cite{cognite_oil_gas}, original equipment manufacturers (OEMs), and shipping companies. Cognite has built a software package named Cognite Data Fusion that extracts useful information from the data. One of the main features of Cognite Data Fusion is its APIs, SDKs, and libraries that are open-source to its customers. Developers and analysts can build applications and ML models that best suit the operation needs. These applications can include large CAD models, complex asset plans, and can be run on phones and tablets. Cognite Data Fusion also offers to customize permissions hierarchies for sharing data with partners and suppliers. For example, the Cognite Digital Platform has helped Framo (an OEM for pumping systems) and their customer Aker BP to communicate and share the live operational data of equipment more efficiently \cite{cognite_OEM}. This enabled Framo to create their applications and monitor the status of equipment to plan the maintenance. Using the operational data of their equipment, OEMs can inform their customers about how to improve the performance of the equipment. The integration of the Siemens information management system (IMS) and Cognite data platform has benefited Aker BP in optimizing the offshore maintenance and reduce costs \cite{cognite_siemens}. With the availability of live data and using artificial intelligence, ML algorithms, Siemens presented a powerful analysis of each equipment with advanced visualizations.     

Siemens digital twin leads the industry by offering diverse computational tools in CAE, CAD, manufacturing and electronic design and connect information from all of these domains using a seamless digital thread to give  companies tremendous insight into products and designs \cite{daveseimens22, davetaylor11}. Siemens plant simulation (PS) digital tool was successfully interfaced (digital copy) with  production line involving manufacturing of pneumatic cylinders within automotive industry to promote the concept of Industry 4.0. The production line simulation model was optimized by genetic algorithm provided by Siemens PS tool  which  adjusted the simulation model and then simulated the digital twin \cite{vachalek2017digital}. Smart factories are designed  with machines that can operate based on manufacturing environments, control production processes and share information with each other in the form of knowledge graphs. Generally, knowledge graphs are incomplete and missing data has to be inferred. Siemens digital twin powered by machine learning tools (e.g., recurrent neural networks) was demonstrated to complete the knowledge graph and synchronize the digital and the physical representations of a smart factory \cite{ringsquandl2017event}.

ANSYS introduced digital twin builder in its ANSYS 19.1 version. 
ANSYS twin builder provides the developer with several features such as creating a multi-domain system, multiple fidelity and multiphysics solver, efficient ROM construction capabilities, third-party tool integration, embedded software integration, as well as system optimization \cite{ansys_dt}. ANSYS along with other companies built the digital twin for a pump that can use the real-time sensor data to improve its performance and to predict failures \cite{ansys_pump}. General Electric also used a customized version of ANSYS digital twin to design megawatt-sized electric circuit breakers \cite{ansys_ge}. 

Akselos, founded in 2012, offers instantaneous physics based simulations and analyses of critical infrastructures calibrated with sensor data in an asset-heavy industry. The key benefits of Akselos digital twin are the asset performance optimization and life extension, failure prediction and prevention, as well as contingency planning. Akselos owns a structural analysis tool that’s fast enough to integrate, re-calibrate, and re-analyze the sensor data. Their framework uses a reduced basis finite element analysis (RB-FEA) technology, a state-of-the-art reduced order modeling approach which is quite faster than conventional FEA and higher accuracy is ensured by using a posteriori accuracy indicators and automated model enrichment. Akselos provides a cloud based platform to develop the digital twin framework for any number of users from any geographic locations or organizations. The structural models developed by Akselos have the capacity to incorporate the localized nonlinearities as well. Among the existing case studies, Akselos digital twin is used for offshore asset life extension, optimizing floating production storage and offloading, inspection, return on investment, and ship loader life extension. Akselos unlocked 20 years of structural capacity for the assets operated by Royal Dutch Shell in the North Sea. Details about the Akselos and their work agenda can be found in \cite{AskelosSA} and \cite{DavidKnezevic2017}.

General Electric (GE) has been developing a digital twin environment integrated with different components of the power plant that takes into account customer defined Key Performance Indicator (KPIs) and business objectives by measuring asset health, wear and performance. Their Digital Twin runs on the Predix™ platform, designed to operate large volumes of sensor data at an industrial scale. Their platform offers advanced distribution management solutions, geospatial network modeling solutions, grid analytics, and asset performance management for power and utility services such as next generation sensing technologies, digital threading, artificial intelligence, advanced control and edge computing. Many world renowned companies have been applied these technologies for diverse industrial fields like automotive, food and beverage, chemicals, digital energy, steel manufacturing, equipment manufacturing, pulp/paper manufacturing, and semiconductors.  Details on GE digital can be found in \cite{gedigital} and \cite{GeneralElectriccompany2016}.

Oracle IoT Cloud offer Digital Twin through three pillars, (i) virtual twin where the physical asset or device is represented virtually in the cloud, (ii) predictive twin using either physics based models (FEM/CFD)  or statistic / ML models having support from Oracle's products such as Oracle R Advanced Analytics for Hadoop (ORAAH) and  Oracle Stream Explorer, and (iii) twin projections where the insights generated by digital twin is projected to the backend application and supported by Oracle ERP (supply chain, manufacturing, maintenance applications) and CX (service) \cite{oracleDT2017}.

\subsection{Human-Machine Interface}
As the demarcation between humans and machine starts to fade in the context of Digital Twin, there will be a need for more effective and fast communication and interaction. While augmented/virtual reality, without doubt will be required to create a detailed visualization of the assets, natural language processing and gesture control will be a very common mode of interaction. We detailed the current state of the art in human machine interface and their potential usage in a DT context.

\subsubsection{Augmented and Virtual Reality}
Augmented Reality (AR) and Virtual Reality (VR) can be considered among the key technologies that promise to add new perspectives in many sectors \cite{yu2009useful}. We refer to several review articles on the state-of-the-arts in engineeing and design \cite{fraga2018review,palmarini2018systematic,rankohi2013review,zhou2008trends,green2008human}, medicine \cite{meola2017augmented,kim2017virtual,barsom2016systematic,berryman2012augmented,sielhorst2008advanced} and education \cite{chen2017review,radu2014augmented,bacca2014augmented,phon2014collaborative}. In \cite{iale}, Escorsa highlighted numerous patented developments within the digital twin context in arrangements for interaction with the human body such as, for example, "haptics" touch feedback technologies as computer generated output to the user developed by Immersion Inc.


\subsubsection{Natural Language Processing}
Voice has been perhaps the most effective and quickest mode of communication among humans. For a seamless integration of the humans and machine in the context of the digital twin, there is an obvious need for elevating this mode of communication to a level that humans and machines can interact seamlessly. With the recent advancements in DL \cite{Goodfellow-et-al-2016} and LSTM algorithms \cite{hochreiter1997long}, language translations and interpretation has reached at least human level accuracy and efficiency if not more. A nice discussion on the major recent advances in Natural Language Processing (NLP) focusing on neural network-based methods can be found in \cite{ruder2019areview}. The blog condenses 18 years (2001-2018) worth of work into eight milestones that are the most relevant today. Further insights about the recent trends in NLP can be found in \cite{young2018recent}.

\subsubsection{Gesture Control}

Recent advancement in remote sensing technologies allows highly accurate gesture recognition capabilities through RF and mmWave radar \cite{he2015wig,lien2016soli,wang2016interacting,li2016wifinger}, ambient light \cite{pathak2015visible,venkatnarayan2018gesture,li2018self}, cameras and image processing \cite{suarez2012hand,jiang2018gesture,zengeler2019hand,mcbride2019comparison}, sound \cite{kalgaonkar2009one,gupta2012soundwave,wang2016device,liu2018toward} and wearable devices \cite{park2011gesture,jung2015wearable,byun2018hand}. Along with the faster communication in the IoT, such technological developments in gesture training and control will be crucial in developing more robust digital twin systems.   

\section{Socio-economic impacts}
\label{sec:socioimpact}
Digital Twinning will bring about unprecedented automation in the management of any physical asset. One of the first concern that can be a stumbling block for the adaptation of digital twins just like any other automation technologies will be its acceptability by the work force. The fear of loosing jobs seemed very logical a few decades earlier \cite{RODD1987285}, however even at that time contrary opinions were prevalent based on systematic studies \cite{MARGULIES1983677}. In fact, Sheridan showed that automation just results in redistribution of workplace without much impact on employment \cite{SHERIDAN1983605}. Such studies highlights the vulnerability of the workforce with lower qualification involved in repeatative jobs. There are also positive aspects with automation and that is with a careful task allocation between humans and machines, it can enable greater safety and creativity at work place by deallocating dirty, dumb and dangerous (3D) jobs to machines and artificial intelligence. Such task allocation has been extensively studied in \cite{DEARDEN2000289, DEVRIES2003719, SHERIDAN2000203}. Humans should be in the loop not only for coordinating AI developments but also checking AI results. At this point it is worth remembering the reflection "ironies of automation" by Lisanne Bainbridge. It simply states that the as automation takes over the simpler works, the role of humans to manage more complicated unpredictable tasks will become even more critical. How to groom humans to deal with unexpected events after prolonged phases of inactivity will be a major challenge. Moreover, for economic reasons, there will be a lesser motivation for recruiting professional users of the relevant technologies, a base of non-professional users might be a natural outcome. This will require strategies to handle security and hacking \cite{CHEN201835}. As the world gears towards greater autonomy resulting from digital twinning, efforts will have to be made to create opportunities for all and nor for a selected few. Nevertheless, with good guiding intentions, the technology will improve the quality of work at work places and with good training and career couselling will enable workers to focus on more creative work.    

\section{Conclusions and Recommendations} 
Based on our literature survey we present the following definition of digital twin: \\
\newline
\vspace{10pt}
\fbox {
	\parbox{0.95\linewidth}{
		\textit{A {\bf digital twin} is defined as a virtual representation of a physical asset enabled through data and simulators for real-time prediction, monitoring, control and optimization of the asset for improved decision making throughout the life cycle of the asset and beyond.
		}
	}
}
We find the breakdown of digital twin into three pillars as proposed by Oracle~\cite{oracleDT2017} to be useful both in  communicating the concept as well as identify the role we think the different stakeholders should take in order to fully exploit its potential:  
\begin{itemize}
	\item {\it Virtual Twin:\/}  Creation of a virtual representation of a physical asset or a device in the cloud.
	\item {\it Predictive Twin:\/} Physics based, data driven or hybrid models operating on the virtual twin to predict the behaviour of the physical asset.
	\item {\it Twin Projection:\/} Integration of insights generated by the predictive twin into the business operation and processes. 
\end{itemize}

Despite the challenges, applying digital twin technologies to various sectors is gaining in popularity. Based on our technology watch, we would like to conclude our analysis providing recommendations for each stakeholders group (Fig.~\ref{fig:stakeholders}) and their most important contributions:

\begin{figure}[htbp]
	\centering
	\includegraphics[width=\linewidth]{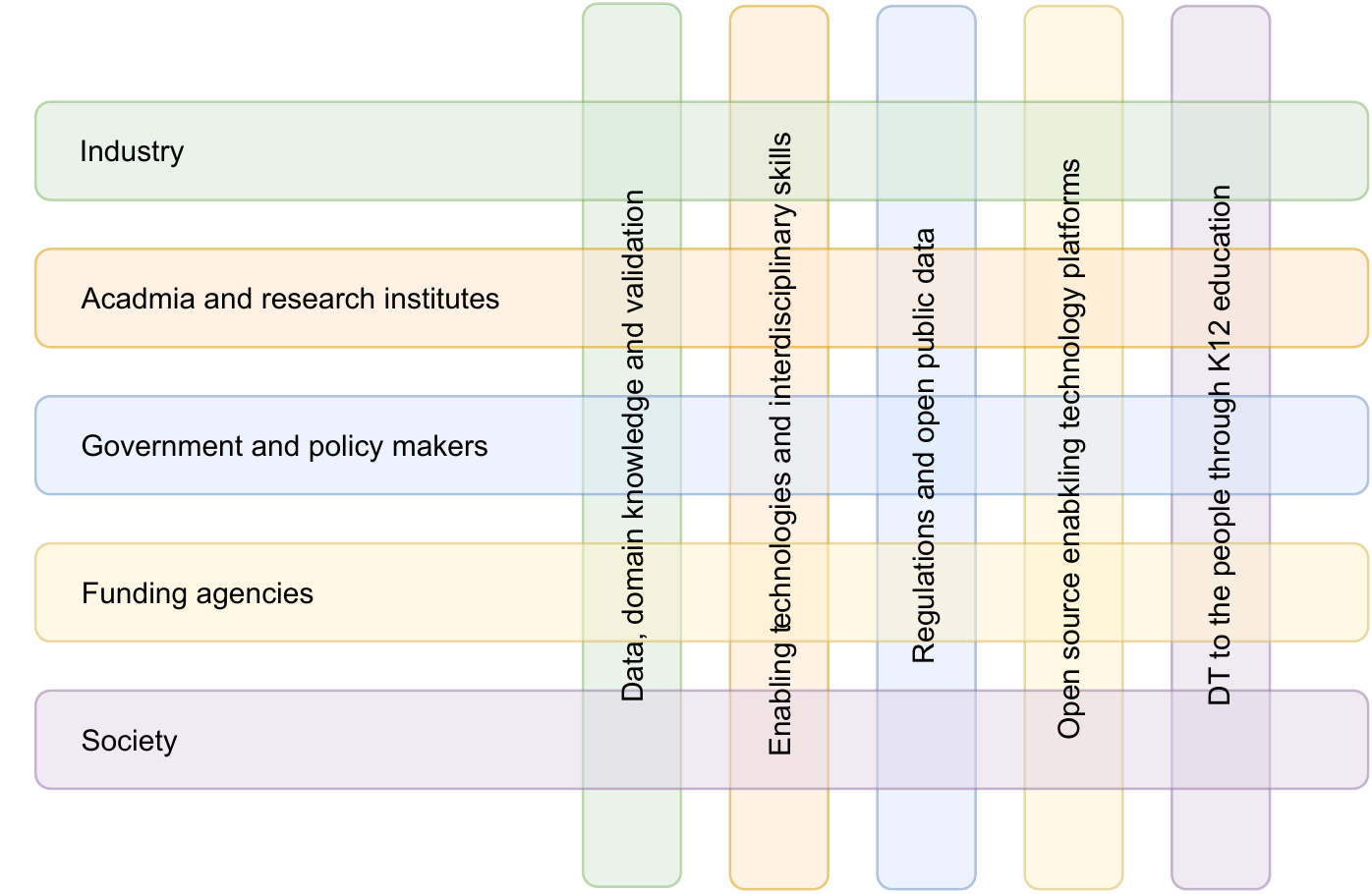}
	\caption{Stakeholders and their potential contributions}
	\label{fig:stakeholders}
\end{figure}

\begin{itemize}
	\item \textit{Industry}: The greatest pull for the technology is going to come from the industry sector. There are three most important ways in which this sector can make positive contributions (a) by making asset dataset available for research and validated model building (b) by actively participating in research through practical knowledge sharing (c) by being proactive in 'projecting' the insights obtained from predictive twins into their business applications to validate the usefulness of digital twins.
	\item \textit{Academia and research institutes}: It is foreseen that along with close cooperation with the industry sectors most of the enabling technologies for both 'virtual twins' and 'predictive twins' will be developed by the academia and research institutes. We strongly recommend that these developments are made exploitable for the society at large  by means of open source software.  Along with providing the technological know-how, the  academia will also have to take the lead in grooming an altogether new generation of workforce to serve the reformed environment. We advocate the MAC-model, i.e.\ to educate and train a new generation of researchers in forming interdisciplinary teams that combine application knowledge (A) with expertise and advanced methodologies from mathematics (M) and computer science (C).  
	\item \textit{Government and policy makers}: To ensure that the benefits of the new technology reaches every layer of the society, that humans are still relevant and there is no compromise with ethics, privacy and security, new inclusive policies and regulations will have to be framed. There should be a major efforts towards the democratization of the technology. Data protection and privacy laws (e.g., the General Data Protection Regulation (GDPR), agreed upon by the European Parliament and Council in April 2016) is definitely and encouraging step in the right direction. The different governmental bodies should initiate feasibility studies for utilizing digital twins within their sectors. In particular, investigate different scenarios on how they can profit from 'projecting' insights obtained from relevant predictive twins into their governmental responsibilities.
	Furthermore, data generated by means of public funding (e.g. weather forecast, anonymous health care data) should be openly available and made easy exploitable for the academia and the industry.
	\item \textit{Funding agencies}: Especially those with a strong mission focusing on industrial innovation impact, the digital twinning offers a challenging theme for center projects (e.g., US NSF Engineering Research Centers and other international comparators \cite{eoin2017}). As a platform for multidisciplinary research, digital twin concept consists of a wide spectrum from the fundamental research and enabling technology development to the system integration phase.
	However, funding of open source enabling technology platforms should be prioritized as the funding mechanisms of such infrastructure have been scarce up to now. 
	\item \textit{Society}: The onus to be well informed about the new technology is on the society itself. It is not a matter of if but when the new technology will bring about remarkable changes in our private and professional lives. Thus, we recommend that one start develop new skills during the K12 education which will facilitate embracement of the emerging technology. 
\end{itemize}

While the definition of a digital twin is unambiguous, what is not a digital twin is a difficult question to answer. There is a need to quantify the degree to which a digital twin resembles its physical counterpart both in terms of appearance and behaviour. As we have discussed in this paper, the focus so far has mostly been on the former but only a little has been done to add physical realism. If we were to define a score of digital twinning on a scale of 1-10, a score of 10 will correspond to a situation when a digital twin becomes indistinguishable from its physical counterpart to an observer. Towards this goal, a hybrid analysis and modeling methodology can be developed by combining ML with physics based models to contribute a firm unified foundation for the computational modeling and simulation of complex problems that arise in numerous multidisciplinary applications.

Finally, we would like to highlight the importance of standardization. It can be argued that in a fully connected and interactive world, different physical assets will be interacting with each other and the corresponding digital twins will also have to interact with each other. In order to facilitate these interactions there will be a need for standards cutting across different domain areas. These standards can range from the file format of the data storage, to the details of how the data can be compressed, to the data protection requirements addressing differences in the laws when spread over different geographical areas. 


To conclude, the digital twin concept offers many new perspectives to our rapidly digitalized society and its seamless interactions with many different fields. Combining complementary strengths in physics-based and data-driven modeling approaches, the hybrid analysis and modeling framework becomes particularly appealing for developing robust digital twin platforms, and enables us to make more informed decisions and ask even better questions to mitigate challenges relevant to big data cybernetics, security, digitalization, automatization and intelligentization.








\section*{Acknowledgment}
The authors would like to thank Professor Harald Martens for his helpful comments and suggestions on an earlier draft of this manuscript.
\bibliographystyle{IEEEtran}
\bibliography{references,references2,references3}
\vskip -2\baselineskip plus -1fil
\begin{IEEEbiography}[{\includegraphics[width=1in,height=1.25in,clip,keepaspectratio]{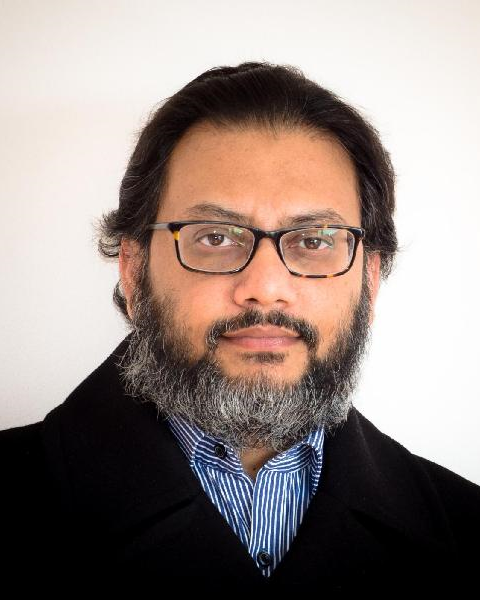}}]{Adil Rasheed} is the professor of Big Data Cybernetics in the Department of Engineering Cybernetics at the Norwegian University of Science and Technology where he is working to develop novel hybrid methods at the intersection of big data, physics driven modelling and data driven modelling in the context of real time automation and control. He also holds a part time senior scientist position in the Department of Mathematics and Cybernetics at SINTEF Digital where he led the Computational Sciences and Engineering group between 2012-2018. He holds a PhD in Multiscale Modeling of Urban Climate from the Swiss Federal Institute of Technology Lausanne. Prior to that he received his bachelors in Mechanical Engineering and a masters in Thermal and Fluids Engineering from the Indian Institute of Technology Bombay.
\end{IEEEbiography}
\vskip -2\baselineskip plus -1fil
\begin{IEEEbiography}[{\includegraphics[width=1in,height=1.25in,clip]{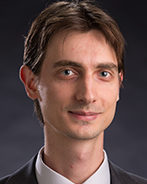}}]{Omer San} received his bachelors in aeronautical engineering from Istanbul Technical University in 2005, his masters in aerospace engineering from Old Dominion University in 2007, and his Ph.D. in engineering mechanics from Virginia Tech in 2012. He worked as a postdoc at Virginia Tech from 2012-'14, and then from 2014-'15 at the University of Notre Dame, Indiana.  He has been an assistant professor of mechanical and aerospace engineering at Oklahoma State University, Stillwater, OK, USA, since 2015. He is a recipient of U.S. Department of Energy 2018 Early Career Research Program Award in Applied Mathematics. His field of study is centered upon the development, analysis and application of advanced computational methods in science and engineering with a particular emphasis on fluid dynamics across a variety of spatial and temporal scales. 
\end{IEEEbiography}
\vskip -2\baselineskip plus -1fil
\begin{IEEEbiography}[{\includegraphics[width=1in,height=1.25in,clip]{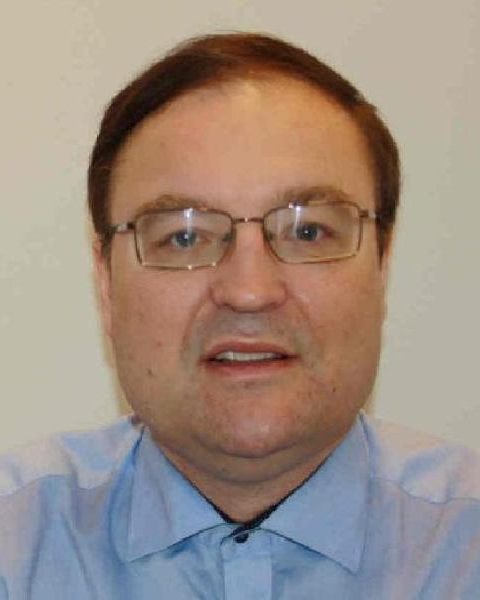}}]{Trond Kvamsdal} is a Professor in the Department of Mathematical Sciences, Faculty of Information Technology and Electrical Engineering at the Norwegian University of Science and Technology. His positions at NTNU are within computational mathematics, i.e. development of new theories/methods within applied mathematics and numerical analysis to make robust and efficient numerical software programs for challenging applications in science and technology. Main area of application is computational mechanics, i.e. both solid/structural and fluid mechanics relevant for civil, mechanical, marine, and petroleum engineering as well as biomechanics, geophysics and renewable energy. He received the IACM Fellow Award (International Association for Computational Mechanics) in 2010 and was elected
	member of the Norwegian Academy of Technological Sciences (NTVA) in 2017. 
\end{IEEEbiography}
\end{document}